\documentclass[11pt]{article}

\usepackage{mathptmx}

\usepackage{graphicx}
\usepackage{latexsym}

\usepackage{pifont}
\usepackage{booktabs}
\usepackage{amsmath,amsfonts,amssymb,nicefrac}
\usepackage{algorithm,algorithmic}
\usepackage{tabularx,multirow}
\usepackage{subfigure,color,epsfig}

\makeatletter%
\newcommand{\singlespacing}{\let\CS=
\@currsize\renewcommand{\baselinestretch}{1}\tiny\CS}
\newcommand{\singlespacingplus}{\let\CS=
\@currsize\renewcommand{\baselinestretch}{1.25}\tiny\CS}
\newcommand{\doublespacing}{\let\CS=
\@currsize\renewcommand{\baselinestretch}{1.75}\tiny\CS}
\newcommand{\extradoublespacing}{\let\CS=
\@currsize\renewcommand{\baselinestretch}{1.9}\tiny\CS}
\newcommand{\nicenicespacing}{\let\CS=
\@currsize\renewcommand{\baselinestretch}{1.9}\tiny\CS}
\newcommand{\draftspacing}{\let\CS=
\@currsize\renewcommand{\baselinestretch}{2.0}\tiny\CS}
\newcommand{\hugedraftspacing}{\let\CS=
\@currsize\renewcommand{\baselinestretch}{2.4}\tiny\CS}
\newcommand{\niceonespacing}{\let\CS=\@currsize\renewcommand{\baselinestretch}{1.1}\tiny\CS}
\newcommand{\nicetwospacing}{\let\CS=\@currsize\renewcommand{\baselinestretch}{1.2}\tiny\CS}
\newcommand{\nicethreespacing}{\let\CS=\@currsize\renewcommand{\baselinestretch}{1.3}\tiny\CS}
\newcommand{\singlespacingplusplus}{\let\CS=\@currsize\renewcommand{\baselinestretch}{1.35}\tiny\CS}
\newcommand{\nicefourspacing}{\let\CS=\@currsize\renewcommand{\baselinestretch}{1.4}\tiny\CS}
\newcommand{\nicefivespacing}{\let\CS=\@currsize\renewcommand{\baselinestretch}{1.5}\tiny\CS}
\newcommand{\nicesixspacing}{\let\CS=\@currsize\renewcommand{\baselinestretch}{1.6}\tiny\CS}
\newcommand{\nicesevenspacing}{\let\CS=\@currsize\renewcommand{\baselinestretch}{1.7}\tiny\CS}
\newcommand{\niceeightspacing}{\let\CS=\@currsize\renewcommand{\baselinestretch}{1.8}\tiny\CS}
\newcommand{\niceninespacing}{\let\CS=\@currsize\renewcommand{\baselinestretch}{1.9}\tiny\CS}
\makeatother%

\newcommand{\normalspacing}{\singlespacing}

\newcommand\qedblob{\mbox{\ding{113}}}

\newcommand{\cale}{{\cal E}}
\newcommand{\condition}{\,{\mbox{\large$|$}}\:}

  \newtheorem{theorem}{Theorem}[section]
  \newtheorem{corollary}[theorem]{Corollary}

\newcommand{\manyonereducesto}{\ensuremath{\leq_{m}^{p}}}

\newcommand{\p}{\ensuremath{\mathrm{P}}}
\newcommand{\np}{\ensuremath{\mathrm{NP}}}
\newcommand{\conp}{\ensuremath{\mathrm{coNP}}}

\newcommand{\pspace}{\ensuremath{\mathrm{PSPACE}}}

\newcommand{\OMIT}[1]{} %

\def\literalqed{{\ \nolinebreak\hfill\mbox{\qedblob\quad}}}

\newenvironment{proofs}{\noindent{\sc Proof.}}{\literalqed\medskip}

\newcommand{\sproofketch}{\noindent{\sc Proof Sketch.}\hspace*{1em}}
\newcommand{\eproof}{\literalqed\medskip}

\newcommand{\qbf}{\ensuremath{\mathrm{QBF}}}
\newcommand{\sat}{\ensuremath{\mathrm{SAT}}}

\newcommand{\score}{\ensuremath{\mathit{score}}}

\newcommand{\ccav}{{\cale\hbox{-}\mathrm{CCAV}}}
\newcommand{\ccdv}{{\cale\hbox{-}\mathrm{CCDV}}}

\newcommand{\ccpvtp}{{\cale\hbox{-}\mathrm{CCPV}}}

\newcommand{\systemccpvtp}[1]{{\mathrm{{#1}}\hbox{-}\mathrm{CCPV}}}

\newcommand{\dcav}{{\cale\hbox{-}\mathrm{DCAV}}}
\newcommand{\dcdv}{{\cale\hbox{-}\mathrm{DCDV}}}

\newcommand{\dcpvtp}{{\cale\hbox{-}\mathrm{DCPV}}}

\newcommand{\systemdcpvtp}[1]{{\mathrm{{#1}}\hbox{-}\mathrm{DCPV}}}
\newcommand{\onlineccac}{{\mathrm{online}\hbox{-}\cale\hbox{-}\mathrm{CCAC}}}

\newcommand{\onlineccdc}{{\mathrm{online}\hbox{-}\cale\hbox{-}\mathrm{CCDC}}}

\newcommand{\onlineccav}{{\mathrm{online}\hbox{-}\cale\hbox{-}\mathrm{CCAV}}}
\newcommand{\onlineccdv}{{\mathrm{online}\hbox{-}\cale\hbox{-}\mathrm{CCDV}}}

\newcommand{\onlineccpvtp}{{\mathrm{online}\hbox{-}\cale\hbox{-}\mathrm{CCPV}}}

\newcommand{\onlinelimitccavcale}[1]{{\mathrm{online}\hbox{-}\cale\hbox{-}\mathrm{CCAV}[#1]}}
\newcommand{\onlinelimitccavcaleprime}[1]{{\mathrm{online}\hbox{-}\cale'\hbox{-}\mathrm{CCAV}[#1]}}
\newcommand{\onlinelimitccdv}[1]{{\mathrm{online}\hbox{-}\cale\hbox{-}\mathrm{CCDV}[#1]}}
\newcommand{\onlinelimitccav}[1]{{\mathrm{online}\hbox{-}\cale\hbox{-}\mathrm{CCAV}[#1]}}

\newcommand{\onlinesystemccac}[1]{{\mathrm{online}\hbox{-}\mathrm{{#1}}\hbox{-}\mathrm{CCAC}}}

\newcommand{\onlinesystemccdc}[1]{{\mathrm{online}\hbox{-}\mathrm{{#1}}\hbox{-}\mathrm{CCDC}}}

\newcommand{\onlinesystemccav}[1]{{\mathrm{online}\hbox{-}\mathrm{{#1}}\hbox{-}\mathrm{CCAV}}}
\newcommand{\onlinesystemccdv}[1]{{\mathrm{online}\hbox{-}\mathrm{{#1}}\hbox{-}\mathrm{CCDV}}}
\newcommand{\onlinesystemccpvte}[1]{{\mathrm{online}\hbox{-}\mathrm{{#1}}\hbox{-}\mathrm{CCPV\hbox{-}TE}}}
\newcommand{\onlinesystemccpvtp}[1]{{\mathrm{online}\hbox{-}\mathrm{{#1}}\hbox{-}\mathrm{CCPV}}}
\newcommand{\onlinedcac}{{\mathrm{online}\hbox{-}\cale\hbox{-}\mathrm{DCAC}}}

\newcommand{\onlinedcdc}{{\mathrm{online}\hbox{-}\cale\hbox{-}\mathrm{DCDC}}}

\newcommand{\onlinedcav}{{\mathrm{online}\hbox{-}\cale\hbox{-}\mathrm{DCAV}}}
\newcommand{\onlinedcdv}{{\mathrm{online}\hbox{-}\cale\hbox{-}\mathrm{DCDV}}}

\newcommand{\onlinedcpvtp}{{\mathrm{online}\hbox{-}\cale\hbox{-}\mathrm{DCPV}}}

\newcommand{\onlinesystemdcac}[1]{{\mathrm{online}\hbox{-}\mathrm{{#1}}\hbox{-}\mathrm{DCAC}}}

\newcommand{\onlinesystemdcdc}[1]{{\mathrm{online}\hbox{-}\mathrm{{#1}}\hbox{-}\mathrm{DCDC}}}

\newcommand{\onlinesystemdcav}[1]{{\mathrm{online}\hbox{-}\mathrm{{#1}}\hbox{-}\mathrm{DCAV}}}
\newcommand{\onlinesystemdcdv}[1]{{\mathrm{online}\hbox{-}\mathrm{{#1}}\hbox{-}\mathrm{DCDV}}}
\newcommand{\onlinesystemdcpvte}[1]{{\mathrm{online}\hbox{-}\mathrm{{#1}}\hbox{-}\mathrm{DCPV\hbox{-}TE}}}
\newcommand{\onlinesystemdcpvtp}[1]{{\mathrm{online}\hbox{-}\mathrm{{#1}}\hbox{-}\mathrm{DCPV}}}

\newcounter{alg}

\newenvironment{algorithmusfall}{\begin{list}
   {{\bf Case~\arabic{alg}:}}
   {\usecounter{alg}}}{\end{list}}

\newcounter{subalg}

\newcommand{\lahnote}[1]{}  \newcommand{\jrnote}[1]{} \newcommand{\ehnote}[1]{}
\newcommand{\LAHnoteHandled}[1]{}  \newcommand{\JRnoteHandled}[1]{} \newcommand{\EHnoteHandled}[1]{}

\newcommand{\pair}[1]{\mathopen\langle{#1}\mathclose\rangle}

\title{The Complexity of Online Voter Control\\in Sequential Elections\thanks{Supported in part by grants
ARC-DP110101792, 
DFG-RO-1202/15-1,
and NSF-CCF-\{0915792,\allowbreak{}1101452,\allowbreak{}1101479\},
and by 
COST Action IC1205,
Friedrich Wilhelm Bessel Research Awards from the 
Alexander von Humboldt Foundation,
the NRW-MIWF 
project 
``Online Partizipation,'' 
and
the SFF grant ``Cooperative Normsetting'' from HHU D{\"u}sseldorf.
Work done
in part while E.~and L.~Hemaspaandra 
were visiting Heinrich-Heine-Universit\"{a}t D\"{u}sseldorf,
and while 
J.~Rothe was visiting the University of Rochester.}}%

\author{Edith Hemaspaandra\\
        Department of Computer Science \\
        Rochester Institute of Technology \\
        Rochester, NY 14623, USA 
\and
        Lane A. Hemaspaandra\\ 
        Department of Computer Science \\
        University of Rochester \\
        Rochester, NY 14627, USA
\and
        J{\"o}rg Rothe\\
        Institut f\"ur Informatik \\
        Heinrich-Heine-Universit{\"a}t D{\"u}sseldorf  \\
        40225 D\"usseldorf, Germany
}
\date{March 2, 2012; revised June 16, 2016} 

\begin{document}
\normalspacing
\sloppy

\maketitle

\begin{abstract}  
  Previous work on voter control, which refers to situations where a
  chair seeks to change the outcome of an election by deleting,
  adding, or partitioning voters, takes for granted that the chair
  knows all the voters' preferences and that all votes are cast
  simultaneously.  However, elections are often held sequentially and
  the chair thus knows only the previously cast votes and not the
  future ones, yet needs to decide instantaneously which control action
  to take.  We introduce a framework that models \emph{online voter
    control in sequential elections}.  
  We show that the related
  problems can be much harder than in the standard (non-online) case:
  For certain election systems, even with efficient winner problems,
  online control by deleting, adding, or partitioning voters is
  $\pspace$-complete, even if there are only two candidates.  In
  addition, we obtain (by a new characterization of coNP in terms of 
  weight-bounded alternating Turing machines) 
   completeness for $\conp$ in the deleting/adding
  cases with a bounded deletion/addition limit, and we obtain 
  completeness for $\np$ in the
  partition cases with an additional restriction.  We also show that for
  plurality, online control by deleting or adding voters is 
  in~$\p$, and for partitioning voters is $\conp$-hard.  
\end{abstract}

\section{Introduction}
\label{sec:introduction}

\lahnote{Joerg:  
ECAI style is that section titles are all upper case, I think...
see their template... so it should I guess be INTRODUCTION and so on 
for other section titles... that looks ugly to me, but that seems
to be their template.  But I don't care much about this... leaving it 
lower case seems fine to me too.  But maybe ECAI people will be 
happier if we follow their approach---I do not know.}%
Elections are important not just in the human world. 
They also can function as an important way of aggregating 
the preferences of 
(often electronic) agents, in our 
world that is increasingly networked and in which 
people and 
institutions 
will increasingly be spoken for by 
automated agents.  

In the field of multiagent systems, voting has been 
suggested for 
tasks as varied as, for example, 
recommender systems,
collaborative spam filtering, and 
planning~\cite{eph-ros:j:multiagent-planning,gho-her-mun-sen:c:voting-for-movies,dwo-kum-nao-siv:c:rank-aggregation}.
And not surprisingly,
the study of the computational properties of voting systems has been
an exceedingly active area within computational social choice.  

In
particular, various types of manipulation, electoral control, and
bribery in voting have been classified in terms of their computational
complexity
(see \cite{fal-hem-hem-rot:b:richer,fal-hem-hem:j:cacm-survey}).
This paper focuses on \emph{voter control}, a model introduced by
Bartholdi, Tovey,
and Trick~\cite{bar-tov-tri:j:control},
where a chair attempts to alter the outcome of an
election via changing its structure by deleting, adding, or partition
of voters.
These types of control seek to model such real-world 
behaviors as targeted vote suppression, bring-out-the-vote
drives, and districting/gerrymandering.
\ehnote{In the abstract we changed partition of voters to partitioning
voters.  For the rest of the paper, we will leave it as partition
of voters.}%
Bartholdi, Tovey,
and Trick's paper was in the bounded-rationality
spirit of Simon~\cite{sim:b:bounded-rationality}, and was
in part making the point that
computational complexity is important in decision-making.

There have been 
many papers analyzing the (non-online) control complexity of 
election systems, and seeking to find natural systems that make 
many types of control attack difficult 
(see the surveys~\cite{fal-hem-hem-rot:b:richer,fal-hem-hem:j:cacm-survey},
the book
chapters~\cite{bau-rot:b:preference-aggregation-by-voting,fal-rot:b:handbook-comsoc-control-and-bribery,hem-hem-rot:b:single-peaked-chapter}, and 
the references therein).
To the best of our knowledge, all previous work on control 
(see, e.g., \cite{bar-tov-tri:j:control,hem-hem-rot:j:destructive-control,fal-hem-hem-rot:j:llull,hem-hem-rot:j:hybrid,erd-now-rot:j:sp-av,erd-fel-rot-sch:j:control-in-bucklin-and-fallback-voting})
\lahnote{That huge list of citations is adding LOTS of space 
to the bibliography.  that is fine for the full version, but i 
doubt we can afford it for the submission.  there are far more 
important uses for the space I suspect.  Ok... I killed a few 
off to save space for the conference.  for the full version
we will eventually make, we can use the longer list that 
currently is commented out.}%
takes for granted that the chair has full knowledge of all the
voters' preferences and that all votes are cast 
simultaneously.\footnote{An exception is a paper 
by Fitzsimmons, Hemaspaandra, and Hemaspaandra~\cite{fit-hem-hem:c:control-manipulation}
that is, regarding their earliest 
appearing versions, more recent than the present paper, and 
studies a mixed model involving both a chair 
and manipulators, in which 
the manipulative voters set their votes 
after actions by the chair.}
However, in many settings voters vote sequentially and the chair's
task in such a setting may often be quite different: Knowing only the already
cast votes but not the future ones, the chair must decide
\emph{online} (i.e., in that moment) whether there exists a control action
that guarantees success, no matter what votes will be cast later on.
We introduce a framework to model \emph{online voter control in
  sequential elections}.  Our approach is inspired by the area of
``online algorithms'' \cite{bor-ely:b:online-algorithms}---algorithms
running and performing computational actions based only on the input
data seen thus far.

In our framework of online voter control, the chair's task stated
above is based on a ``maxi-min'' idea
(although here, due to the time effects, that
can involve more than two quantifiers), a typical online-algorithmic
theme; in that framing of the chair's task we are 
following the approach that has been used for online 
manipulation and online 
candidate control~\cite{hem-hem-rot:j:online-manipulation,hem-hem-rot:c:online-candidate-control}.  
Note that another central online-algorithmic theme,
a strictly numerical ratio approach to so-called
``competitive analysis,'' would not apply very naturally here; the 
reason is that in its general setting, voting (in social-choice theory) 
is most typically based on an ordinal notion of preferences, and those don't 
convey cardinal strength-of-affinity information regarding the 
outcome. 
(For some specific voting systems such as so-called 
scoring systems one can interpret them as giving cardinal information,
and we commend as an interesting open issue a future, general 
control-complexity study 
for such systems 
in 
terms of a competitive-ratio analysis; see 
\cite{luc-ore:c:budgeted-social-choice}, 
which takes that approach for 
the issue of 
selecting a bundle of goods.)
Sequential (or otherwise ``dynamic'') voting has
been studied in other contexts as well, e.g., from a game-theoretic
perspective as ``Stackelberg voting
games''~\cite{con-xia:c:stackelberg-sequential} (see
also~\cite{des-elk:c:sequential-voting,dek-pic:j:sequential-voting-binary-elections,slo:j:sequential-voting}),
or using an
axiomatic approach~\cite{ten:c:transitive-voting} or
Markov decision processes~\cite{par-pro:c:dynamic-social-choice}.
None of this work has considered the issue of voter control.

What our results show is that such online control problems 
can be much harder than in the standard (non-online) case.
We show that 
for certain election systems, even with efficient winner problems,
  online control by deleting, adding, or partitioning voters is
  $\pspace$-complete, even if there are only two candidates.  
In
  addition, we obtain completeness for $\conp$ in the deleting/adding
  cases with a bounded deletion/addition limit.  We do this 
by establishing a complexity-theoretic result 
(Theorem~\ref{thm:k-dour-QBF}) that is 
of interest in its own right: Polynomial-time 
alternating 
Turing machines that on each accepting path
make a constant number 
of 
``Yes'' guesses 
accept only coNP languages, and in fact this 
completely characterizes $\conp$.
We also show that for
  plurality, online control by deleting or adding voters is 
  in~$\p$, and for partitioning voters is $\conp$-hard.

\section{Motivation}\label{s:motivation}
The coming sections will give our definitions, results, and 
proofs.
However, before that,
the present section will very informally present some motivation
and examples.  In particular, we give example settings in which it is
natural to study sequential action, in which the election's ``chair''
has a use-it-or-lose-it ability to do
addition/deletion/partition-choice for each voter as the voter votes,
and the chair knows the votes of the voters seen so far, but not of
future voters.  Of course, theoretical models don't capture the many
interactions and subtleties of the real world, and so 
our models don't perfectly capture the full richness of 
even these sample situations.
Nonetheless, we feel
that for many cases, such as those we are about to mention, the 
theoretical models
we develop in this paper are far closer to capturing the
real-world situation than are existing models of simultaneous
voting or even existing models where votes are sequential but all
voters' preferences are known ahead of time.

As a concrete example (and let us for the moment not worry about what
the particular election system is), consider a College faculty meeting
at which, going right around the room, the faculty members hand their
handwritten paper ballots to the Dean, who then passes them on to her
administrative assistant, who quietly adds them to the totals he is
keeping.  But let us further assume that the Dean is a shifty person,
and can, for a certain number of ballots, slip the piece of paper into
her pocket after reading the vote, without that being noticed, and
without the people in the room being likely to notice that there
aren't quite enough votes in the totals (let's suppose it is a big
college).  And the question 
is, given that we are at some particular point in going around 
the table (and know what votes have been cast so far and what 
actions the Dean---or whoever was standing in for her---has taken
so far): Can 
the Dean ensure,
using at most her remaining amount of vote-to-pocket shifting,
that the winner(s) of
the election will 
include at least one of 
the  alternatives she favors?
This
setting loosely corresponds to our sequential version of control by
deleting voters.  For vividness, our examples are about 
humans voting and a human chair (in the above, the Dean),
and in the case just given, paper ballots.
However, 
our 
model applies also
to more electronically focused
cases of preference 
aggregation, e.g., the ``Dean'' in the above example
could be a doctored voting machine that can only suppress so many ballots 
before seriously risking detection.

The above example is about deleting voters, but there are also 
natural examples for 
adding or partitioning voters.  For partitioning voters, imagine that
a school's undergraduate admissions office is going to use a panel,
whose members will each be assigned to one of two faculty committees,
to vet applying students (perhaps with the committees purportedly
looking for different things, e.g., one is looking for traditional
smartness and the other is looking for unusual levels of passion and
creativity), with all applicants' folders given to both committees,
and with each committee using voting to select its favorite proposals,
and then with only the winners of those two vetting elections moving on to a
final election in which all the panel members vote.  Suppose a
particular admissions office staff member (who is the chair
in this example), with all the faculty
members lined up and coming into the room, as each faculty member
steps to the doorway briefly chats with the faculty member well enough
to discern the likely votes he or she will cast, and then right there
assigns the person to either the smartness committee or the passion
committee.  If the admissions staff member
does so with the goal of
ensuring that at least one of a certain set of students (perhaps the
students who are great football quarterbacks, or the students whose
parents might fund a new admissions building) will be admitted, that
very loosely put would be captured by our sequential version of
control by partition of voters.\footnote{Actually, 
as our previous example suggested, our model is a bit more 
flexible and allows one to ask such questions starting 
at an intermediate point at which some actions have already 
been taken, potentially by a different
admissions staff 
member.}
For adding voters, a natural model
might be a political candidate (who is the chair in this example) 
going door to door through her district
in a preset order, and knowing from public records which voters are
registered voters and which are not, and at each door meeting and
learning the voter's preferences among the candidates, and then for
those voters who are not registered deciding whether to use charisma
to convince them to register and vote, with the limitation that the
candidate has only so much charisma to use.

The above are a few very informal examples of settings where
sequential action is natural, and one knows the votes cast so far but
not those to be cast in the future (except who will be casting them
and in which order).  Let us finish this informal section by briefly
giving a mini-example of the flavor of the goal we have for our
chairs, and how that affects their actions.  We are assuming that
chairs are very pessimistic: What a chair wants to know is whether there
is some action she can take at the given moment so that one of her
preferred candidates will win no matter what the value is of all the
currently unknown-to-her future votes---but assuming that her own
future decisions are (of course)
aimed at 
supporting her goal.
To make this
more concrete, let us discuss the most important real-world election
system: plurality.  In our addition-of-voters example above, suppose
the candidate going door to door has only one preferred candidate in
the election, namely, herself.  Then it is quite clear and simple what
she should do.  Until she runs out of charisma, she should for each
unregistered voter she meets for whom she is the favorite candidate
expend her charisma to have that person become a registered voter.
That is an \label{word:operational}``operational'' approach 
that would work perfectly.  But
more must be said.  The question our pessimistic candidate (and our
decision problems) wants answered at each point is whether, whatever the
preferences still to come after the current point are, that candidate will
win.  And it is also clear how to judge that.  The candidate, as she
starts speaking with a given unregistered voter who likes our
candidate the most (we can similarly describe how to reason in the
other cases), reasons as follows: I need to assume that all future
voters (whose preferences I don't currently know!)~concentrate their
votes on a candidate other than me who currently has the most votes
(in the tally I have been building in my canvas so far), and that I
use my charisma to (if it is not expended) add the current
unregistered voter, and then I suppress
those hypothetical, unregistered, against-me voters, 
and would that leave me a winner
of this election?  If the answer is yes, then the candidate should be
very happy, as she knows she can guarantee herself victory as long as
she doesn't later do anything overtly stupid with her charisma.  The
example we just gave is in effect explaining why it holds that
(so-called constructive) control by adding voters is in
polynomial time for sequential plurality elections.  Now, one might
assume that plurality is such a simple system that for all types of
sequential control we will obtain polynomial-time algorithms.
However, as Theorem~\ref{thm:plurality-partitioning} we will show that
that is not the case (unless $\rm P= NP$).  The proof of that result
is in effect giving an example---although admittedly a more complex
one---in which a coNP-hard problem, namely the complement of Hitting
Set, is transformed into an election control instance about
sequentially partitioning voters (the control setting we described
above in our example about college admissions).

\section{Preliminaries}
\label{sec:preliminaries}

We assume familiarity with 
standard complexity-theoretic notions
such as the
complexity classes~$\p$, $\np$, $\conp$, and $\pspace$, 
polynomial-time many-one reductions ($\manyonereducesto$), 
$\manyonereducesto$-hardness, and 
$\manyonereducesto$-completeness~\cite{hop-ull:b:automata,pap:b:complexity}.  
A standard $\np$-complete problem is the
satisfiability problem ($\sat$) from propositional logic, a standard
$\conp$-complete problem is the tautology problem, and the quantified
boolean formula problem ($\qbf$) is a standard $\pspace$-complete
problem.

This paper provides both polynomial-time algorithms 
and NP-completeness results.  The latter 
are worst-case results, and so it is possible 
that for certain distributions heuristics might do well
(see~\cite{rot-sch:j:typical-case-challenges} for a survey of 
this in the context of elections).  We commend this direction 
as an area for future research.  However, such studies are 
quite dependent on distributions, and by relatively recent
work,
it is known that 
for the uniform distribution 
heuristic algorithms cannot asymptotically have subexponential 
error frequency on any NP-hard problem unless the polynomial
hierarchy collapses to (and indeed, slightly further than) its 
third level%
{}~\cite{buh-hit:c:np-hard-sets-and-density,cai-cha-hem-ogi:j:s2-and-lowness,hem-wil:j:heuristic-algorithms-correctness-frequency}.
(Note: An algorithm is said to have 
{subexponential error frequency} 
if for 
every $\epsilon > 0$ the number of errors the algorithm makes at 
length $n$ is $O(2^{n^\epsilon})$; see 
\cite{hem-wil:j:heuristic-algorithms-correctness-frequency} for a more 
detailed explanation.)

\subsection{Voter Control Types in Simultaneous Elections}

A pair $(C,V)$ is called a \emph{(standard or simultaneous) election}
if $C$ is a set of candidates and $V$ a list of voters that all have
\ehnote{I am not a big fan of defining voters as lists, but
this is not the time to change it.}%
cast their votes simultaneously.  We assume that each vote in $V$ has
the form $(v,p)$, where $v$ is the name of this voter and $p$ is $v$'s
(total) preference order over~$C$.  For example, if $C = \{c,d,e\}$
then $(\mathrm{Bob},d>e>c) \in V$ indicates that Bob (strictly)
prefers $d$ to $e$ and $e$ to~$c$ (or, to be more precise,
it indicates that that is the ballot Bob has cast).

The
standard types of (constructive) voter control in simultaneous
elections 
are as follows.  (These are as introduced 
by Bartholdi, Tovey, and Trick~\cite{bar-tov-tri:j:control}, except 
here we will follow the now more standard model---called
the nonunique-winner model---of asking whether a 
candidate can be made \emph{a} winner, rather than their
approach---called the unique-winner model---of asking 
whether a candidate can be made the one and 
only winner.)
An election system is a mapping from elections (votes/candidates)
to a winner set.
Let $\cale$ be a given election system.  In
\emph{control by deleting voters} ($\ccdv$), given an election
$(C,V)$, a distinguished candidate $c \in C$, and a nonnegative
integer $k \leq \|V\|$, we ask whether there exists a set 
of at most $k$
voters from $V$ such that $c$ is an $\cale$ winner of the 
election in which that set of
voters is removed.  In \emph{control by adding voters} ($\ccav$), we are given
a candidate set~$C$, a list $V$ of registered voters with preferences
over~$C$, a list $V'$ of as yet unregistered voters with preferences
over~$C$, a distinguished candidate $c \in C$, and a nonnegative
integer $k \leq \|V'\|$, and the question is whether 
there exists a set of 
at most $k$ voters from $V'$ such that $c$ is an $\cale$ winner of the
election where the voters are that set and all of $V$.
Finally, in \emph{control by partition of
  voters}, we are given an election $(C,V)$ and a distinguished
candidate $c \in C$, and we ask whether $V$ can be partitioned into
two sublists, $V_1$ and~$V_2$, such that $c$ is an $\cale$ winner of
the election 
$(W_1 \cup W_2,V)$, 
where $W_i$ for $i \in \{1,2\}$ is
the (possibly empty) set of winners of subelection $(C,V_i)$ that have
survived the tie-handling rule used and by $V$ here we implicitly mean $V$ 
masked down to just those candidates in $W_1 \cup W_2$.  
Of the two tie-handling models
introduced by 
Hemaspaandra, Hemaspaandra, and Rothe~\cite{hem-hem-rot:j:destructive-control}
we focus on the
\emph{ties-promote (TP)} model only, where all winners of a
subelection proceed to the runoff, since that model fits more naturally
with the nonunique-winner model in which we will define our online control
problems.  The resulting problem is denoted by $\ccpvtp$.

The destructive variants of these three problems, denoted by $\dcdv$,
$\dcav$, and $\dcpvtp$, are obtained by requiring that the
distinguished candidate $c$ is \emph{not} a winner of the election
resulting from the control action at
hand~\cite{hem-hem-rot:j:destructive-control}.  

\subsection{Online Voter Control in Sequential Elections}

We study \emph{online voter control in sequential elections}, where we
assume that the voters vote in order, one after the other, 
each expressing preferences over all the candidates.
If $u$ is the current voter and $C$ the given candidate set, an
\emph{election snapshot for $C$ and $u$} 
is specified by a triple $V =
(V_{<u}, u, V_{u<})$, where the earlier voters $V_{<u}$ have already cast
their votes, each a preference order over~$C$, and now it is $u$'s turn to
cast a vote, and the future voters $V_{u<}$ will cast their votes
in the order listed.
($V_{<u}$ and $u$ of course list the votes 
cast 
and who cast them, but 
$V_{u<}$ just gives the order of the voters following $u$.)
This snapshot approach is natural for studying online attacks on 
elections,
and was used previously to study the different type of 
attack known as online manipulation in sequential
elections~\cite{hem-hem-rot:cOutByJour:online-manipulation,hem-hem-rot:j:online-manipulation}.

We now define our notions of online voter control for the standard
voter control types stated above, and the related problems.  They all
will start from a basic \emph{online voter control setting} (an
\emph{OVCS}, for short), augmented by appropriate additional
information according to the control type at hand.  A basic OVCS $(C,
u, V, \sigma, d)$ consists of a set $C$ of candidates, the current
voter~$u$ (which isn't strictly needed here, as $u$ is clearly
singled out within $V$ anyway), 
an election snapshot $V$ for $C$ and~$u$, the chair's
preference order $\sigma$ on~$C$, and a distinguished candidate $d \in
C$.  Let $\cale$ be a given election system and let $W_{\cale}(C,V)$
denote the $\cale$ winner set of (standard) election $(C,V)$.  For
each online voter control type we will define, the question the chair
faces is: 
Does there exist a control-action choice
of our considered type
regarding the current 
voter
(e.g., whether or not to
delete~$u$) such that if the chair takes 
that action, then no matter what votes the remaining voters after
$u$ cast, the chair's goal can be reached by the current
decision regarding $u$ and by using the chair's future decisions 
(if any), each being made using the chair's then-in-hand
knowledge about what votes have been cast
by then?\footnote{Note that this maxi-min-inspired (but with 
more quantifiers) approach 
is really about alternating quantifiers.  We are asking
if there exists a current action of the chair, such that for all 
potential revealed vote
values that come between now and the next time the chair has to 
decide on an action, there exists a next action by the chair, such 
that for all
$\ldots$~$\ldots$~the 
chair reaches her goal.}
By \emph{the
  chair's goal} we mean to ensure $W_{\cale}(C,V') \cap \{c
\condition c \geq_{\sigma} d\} \neq \emptyset$ for each possible
ultimate election $(C,V')$ (i.e., 
each $V'$ is a 
possible vote list
resulting from the control type at hand after all decisions have been
made by the chair and all voters have cast their votes) in the
constructive case, and to ensure that $W_{\cale}(C,V') \cap \{c
\condition d \geq_{\sigma} c\} = \emptyset$ in the destructive
case (i.e., that neither $d$ nor any candidate the chair
likes even less than $d$ is a winner).\footnote{\label{f:why-order}Why do we 
provide an ordering $\sigma$ rather than
just providing as a list the set of candidates who are good enough 
to count as reaching our goal?  For the decision-problem version
of online voter control, which is our formulation here, providing 
such a set would be just as good.  But by making $\sigma$ a part of the 
input, we make the model compatible, for the future, with the interesting 
optimization problem 
of trying to find the most preferred candidate within $\sigma$ for 
which the chair can ensure that there is among the winner set one of the 
candidates in the segment from that candidate to the top candidate in 
$\sigma$.

Also, to avoid any confusion, we note that in our ``$d$ chooses an
upper (constructive case) or lower (destructive case) segment of the
candidates'' approach, the non-online version's situation that the
destructive goal ``opposing'' a constructive goal is specified in the
same way not longer holds (although we could have defined things in a
less natural way so that that would hold).  That is, in the non-online
setting, the distinguished candidate $d$ in the constructive case is
saying who the chair wants to win, and in the destructive case is
saying who the chair wants to not win; $d$ in one case is defined 
in the problem definition to denote the beloved candidate and in the 
other case is defined to denote the despised candidate.  
However, in our case, we are giving an order $\sigma$, and it would be 
perverse and confusing to have $>$ mean one thing for constructive and 
another for destructive.  And so, as we have defined things, 
if the
chair's stated ordering $\sigma$
is $v_1 > v_2 > v_3 > v_4 > v_5$ and $d=v_2$,
in the constructive case that means that the chair wants at least one
of $v_1$ or $v_2$ to win.  To state the destructive-case goal---which
in some sense is the ``flip'' of that constructive-case goal---of
having neither $v_1$ nor $v_2$ be a winner, one would give as the
chair's ordering $v_5 > v_4 > v_3 > v_2 > v_1$ and $d=v_2$, since this 
specifies that $v_2$ and $v_1$ are the chair's 
two most despised candidates
and are the ones the chair wants to prevent from being winners.

These comments simply refer to
the way various ``opposite'' goals happen to be expressed.
None of the above is saying that 
the constructive \emph{problem} (viewed as a set) and the 
destructive \emph{problem} (viewed as a set) 
are each other's complements.  Due to
the quantification involved regarding the actions being taken such as 
by the chair, that is not true.}
Note that the conditions 
$W_{\cale}(C,V') \cap
\{c \condition c \geq_{\sigma} d\} \neq \emptyset$ and
$W_{\cale}(C,V') \cap \{c \condition d \geq_{\sigma} c\} = \emptyset$
defining the chair's goal
have the flavor, give or take the fact that we are focusing 
on a top segment of $\sigma$, of the nonunique-winner model,
e.g., as long as 
$W_{\cale}(C,V') \cap
\{c \condition c \geq_{\sigma} d\} \neq \emptyset$ we call it success
even if more than one candidate ties as winner.
To formally define our problems, it remains to specify for each
control type the information by which the basic OVCS is augmented.
What kind of decisions the chair is to make in the course of a
sequential election will always be clear from the control type at hand
(e.g., whether or not to delete a voter in ``online control by
deleting voters'').

Let $B = (C, u, V, \sigma, d)$ be a given basic {OVCS}.  For
\emph{online control by deleting voters}, $B$ is augmented by the
following additional information: A nonnegative integer $k$ (the
deletion upper bound); for each voter $v$ before~$u$, there is a flag
saying whether $v$ was deleted and the vote cast by $v$ (if not
deleted)---at most $k$ voters can be marked as deleted for the input
to be syntactically legal; and the vote 
the current voter $u$ will cast
(if not selected for deletion).  We denote these problems by
$\onlineccdv$ (constructive) and $\onlinedcdv$ (destructive).
(We certainly could equivalently formulate the problem in a way that
masks out all earlier deleted voters, and so removes the need for the
flagging; but we prefer the above version since it allows the
\label{words:actual-history}actual history of 
the voting situation to be part of the instance.)

For \emph{online control by adding voters}, $B$ is augmented by the
following additional information: A nonnegative integer $k$ (the
addition upper bound); each voter $v$ in $V$ has a flag saying if $v$
is unregistered (i.e., can be added) or registered---$u$ must be
unregistered; each unregistered 
voter $v$ before $u$ has another flag saying if $v$
was added---at most $k$ voters may have that flag set in any syntactically
legal input; the vote cast is given for each registered or added
unregistered voter before~$u$; and also given is 
the vote $u$ will cast (if it is added).
We denote these problems by $\onlineccav$
(constructive) and $\onlinedcav$ (destructive).

For \emph{online control by partition of voters}, $B$ is augmented by
the following additional information: Each voter $v$ before $u$ has a
flag saying which part of the partition $v$ was assigned to (``left''
or ``right'') and the vote cast by~$v$, and also $u$'s vote is given.
We denote these problems by $\onlineccpvtp$ (constructive) and
$\onlinedcpvtp$ (destructive).
As a reminder, the two preliminary elections are conducted 
under the convention that 
``ties promote'' (i.e., all winners of the preliminary elections move
forward to the final election).

A natural worry about our maxi-min approach to online voter control is
that it is always possible that all the future voters are hostile to
one's goals.  And in that case, one may be, depending on the election
system, powerless to reach one's goal in the worst case, and so the
maxi-min outcome is easily seen to be failure to reach one's goal.
Although this worry exists in a weaker form for
\ehnote{Used to say: ``Although this worry exists in a weaker form for
two other issues we are studying,'' but that gives away our identity.}%
online manipulation
and online bribery,
since for those
if one is allowed almost no vote-changing one is in many cases
obviously in trouble, at least in those settings one can do whatever
one wants to those votes one does manipulate or bribe.  In control,
however, one doesn't get to set the value of a single vote, and that is pretty
extreme.

This is a valid worry, but there are some things that keep it in
perspective.  Primarily, our paper is trying to find out the very
greatest complexity that online control in sequential elections can
ever have (when restricted to
election systems having polynomial-time winner problems).  And so we
can look at election systems that sidestep the above worry, due to
their properties simply not matching the intuition above (which assumed
that we are using an election system in which having a lot of bad-for-us
votes results in a bad-for-us outcome).  In effect, we are seeking to
understand the limits of behavior, in order to set a bounding box on
the behaviors that can be realized.  Of course, for many natural
election systems, the effect mentioned in the previous paragraph will
hold, and for many inputs that fact can be exploited to help achieve
polynomial-time algorithms for the control problem; indeed, in this
paper itself, we give examples of achieving polynomial-time algorithms
for the most important of election systems: plurality.
Of course, problems may start with some votes already cast, and this
may itself 
make for interesting ``endgame'' decision
issues.
We also
very much hope further studies will
be conducted 
employing a range of models,
including ones beyond maxi-min.

\section{General Upper and Lower Bounds}

\begin{theorem}
\label{thm:general-pspace-upper-bound}  
For each election system $\cale$ with a polynomial-time winner
problem,\footnote{The statement of
  Theorem~\ref{thm:general-pspace-upper-bound} holds even for election
  systems whose winner problems are in $\pspace$.}
$\onlineccdv$, $\onlinedcdv$, $\onlineccav$, $\onlinedcav$,
$\onlineccpvtp$, and $\onlinedcpvtp$ are in $\pspace$.
\end{theorem}

\begin{proofs}
  The upper bounds follow from the observation that each of these
  problems can be solved by an alternating Turing machine in
  polynomial time, and thus by a deterministic polynomial-space Turing
  machine, by the characterization due to Chandra, Kozen, and 
  Stockmeyer~\cite{cha-koz-sto:j:alternation}.~\end{proofs}

Theorem~\ref{thm:general-pspace-upper-bound}  
settles all general (i.e., regarding any voting system for which
winner determination is easy) upper bounds for our online voter
control problems.  We now turn to exploring their lower bounds.

\subsection{Control by Deleting and by Adding  Voters}

\begin{theorem}
\label{thm:general-pspace-lower-bound-deleting-adding-voters}  
There exist election systems $\cale$ and $\cale'$ with polynomial-time
winner problems such that $\onlineccdv$, $\onlineccav$,
$\onlinesystemdcdv{\cale'}$, and $\onlinesystemdcav{\cale'}$ are
$\pspace$-complete, even when limited to two candidates.
\end{theorem}

\begin{proofs}
Let $(C,V)$ be an election.
\ehnote{Was:  For constructive control by deleting voters, let $(C,u,V,\sigma,d)$
 be a given basic OVCS, augmented by the additional information of
 online control by deleting voters.  But winner problems take
an election as input; I moved the OVCS stuff to where it's used.}%
We define election system
  $\cale$ as follows.  $\cale$ interprets---in some fixed, natural
  encoding---the lexicographically least candidate name in $C$ as a
  boolean formula, $\Phi$,
  whose variable names must be the strings $x_1, x_2, \ldots ,
  x_{2\ell}$ for some~$\ell$, where $x_{2\ell}$ actually appears in
  $\Phi$ (the other variables don't have to; no variables other than
  $x_1, x_2, \ldots , x_{2\ell}$ are allowed).  If these syntactic
  requirements fail to hold, everyone loses in~$\cale$.  Otherwise, if
  any two voters in $V$ have the same name, everyone loses in~$\cale$.
  Otherwise, order the voters in $V$ lexicographically by name of the
  voter, and let $v_1, v_2, \ldots , v_z$ be the voter names
\ehnote{Was: (each a binary string), but we did not define names that way
in this paper}%
  in this order.  If $z < 2\ell$ or if there are less
  than two candidates, everyone loses in~$\cale$.  Otherwise, if for
  some odd~$i$, $1 \leq i \leq 2\ell - 1$, the two lowest order bits
  of $v_i$ are not $00$ or $01$, or if for some even~$i$, $2 \leq i
  \leq 2\ell$, the two lowest order bits of $v_i$ are not $10$ or
  $11$, everyone loses in~$\cale$.  Otherwise, assign the variables of
  $\Phi(x_1, x_2, \ldots , x_{2\ell})$ as follows.  For each odd~$i$,
  $1 \leq i \leq 2\ell - 1$, set $x_i$ to \emph{true} if the two
  lowest order bits of $v_i$ are
\ehnote{was: ``are not''}%
 $01$, and set $x_i$ to
  \emph{false} otherwise (i.e., the two lowest order bits of $v_i$ are
  $00$).  For each even~$i$, $2 \leq i \leq 2\ell$, set $x_i$ to
  \emph{true} if the name of the least preferred candidate in the vote
  of $v_i$ is lexicographically less than the name of the next to
  least preferred candidate in the vote of~$v_i$, and set $x_i$ to
  \emph{false} otherwise.  If this assignment satisfies $\Phi$,
  everyone wins in~$\cale$, and otherwise everyone loses.  This ends
  the specification of~$\cale$.  Since a boolean formula whose
  variables have all been assigned can be evaluated in polynomial
  time, $\cale$ has a polynomial-time winner problem.

  By Theorem~\ref{thm:general-pspace-upper-bound}, $\onlineccdv$ is in
  $\pspace$.  To show $\pspace$-hardness of $\onlineccdv$, we
  $\manyonereducesto$-reduce the $\pspace$-complete problem $\qbf'$, a
  variant of $\qbf$, to it.  $\qbf'$ is the set of
  \jrnote{I'm diverging here from the notes on p.~156 a bit, by adding
    quantifiers}%
  \lahnote{His diverging seems fine and needed I'd guess.  Wait... 
I just looked at page 156 and what is there is fine so 
diverging was not needed... it is framed
on page 156 so as not to need the quantifiers.  But Joerg's decision
to put them in probably is fine too as long as he adjusted 
everything to match that... but i leave that to 
you, Edith to check... if he is diverging, do make sure he isn't
doing something wrong.}%
\ehnote{I undid the diverging.  I have also tried to make it
more clear that we are talking about the actual names of variable.}%
boolean formulas of the form 
$F(x_1, x_2, \ldots , x_{2\ell})$, for some~$\ell$,
  such that the variable $x_{2\ell}$ appears in~$F$, all variables
  appearing in $F$ are from the variable name collection ``$x_1$'',
``$x_2$'', \ldots , ``$x_{2\ell}$'', and
\[
(\exists b_1)\, (\forall b_2)\, \cdots
  (\exists b_{2\ell - 1})\, (\forall b_{2\ell})\, [F(x_1:=b_1,
  x_2:=b_2, \ldots , x_{2\ell}:=b_{2\ell}) \text{ evaluates to
    \emph{true}}],
\]
  where $b_i \in \{0,1\}$ and $x_i:=b_i$ means that
  variable $x_i$ is set to \emph{true} if $b_i=1$, and is set to
  \emph{false} if $b_i=0$, for $1 \leq i \leq 2\ell$.

Let $F(x_1, x_2, \ldots , x_{2\ell})$ be
  a given instance of $\qbf'$, where $x_{2\ell}$ explicitly appears
  in~$F$.  (If our input is syntactically incorrect, we map it to
  a fixed no-instance of $\onlineccdv$.)  We construct
  from $F$ an instance of $\onlineccdv$, consisting of a basic OVCS
$(C,u,V,\sigma,d)$, augmented by the additional information of
online control by deleting voters, as follows.
  Define $C = \{a,b\}$, where $a$ encodes $F$ (in our fixed, natural
  encoding of boolean formulas) and $b$ is the string
  lexicographically immediately following $a$; the current voter is
  $u=v_1$; $V$ will be specified below; the chair's preference order
  is $a >_{\sigma} b$; for specificity, we let $d=a$ be the
  distinguished candidate (though that does not matter, as all
  candidates win or all lose in~$\cale$); the deletion limit is $k =
  \ell$; and a vote $a>b$ to cast for $u$ if not deleted (again, the
  vote doesn't matter, as $u=v_1$ will specify an assignment to
  $x_1$ by her name, not by her vote).  There are
  $\left(\nicefrac{3}{2}\right)\cdot 2\ell = 3\ell$ voters in $V$ such
  that the name of the
  $i$th voter, $v_i$, is the binary string $u_i
  w_i$, where $u_i$ is the binary representation of $i$ and $w_i=00$
  if $i \equiv 1 \bmod 3$, $w_i=01$ if $i \equiv 2 \bmod 3$, and
  $w_i=10$ if $i \equiv 0 \bmod 3$, $1 \leq i \leq 3\ell$.  This
  completes the description of our $\manyonereducesto$-reduction from
  $\qbf'$ to $\onlineccdv$, which clearly can be computed
  in polynomial time.

  We claim that $F \in \qbf'$ if and only if the chair's goal can be
  reached by at most $k$ deletions of voters.  Why?  By the definition
  of~$\cale$, everyone loses unless our $k=\ell$ deletions are used on
  \emph{exactly one of $v_{3i-2}$ and $v_{3i-1}$}, for each~$i$, $1
  \leq i \leq \ell$.  No $v_{3i}$, $1 \leq i \leq \ell$, can be
  deleted if there is to be a winner.  And the ``exactly one of
  $v_{3i-2}$ and $v_{3i-1}$'' choices, $1 \leq i \leq \ell$, specify
  an assignment of truth values to the odd-numbered variables:
  For each~$i$, $1 \leq i \leq \ell$, $x_{2i - 1}$ is set to
  \emph{true} if $v_{3i - 2}$ is deleted and $v_{3i - 1}$ is not, and
  is set to \emph{false} if $v_{3i - 1}$ is deleted and $v_{3i - 2}$
  is not.  On the other hand, for each $i$, $1 \leq i \leq \ell$,
  the truth value of $x_{2i}$ is specified
  by the \emph{vote} of voter $v_{3i}$, since after these $\ell$ deletions,
  $v_{3i}$ will be the 
$2i$th 
voter name in the lexicographic order. 
\ehnote{This was:  On the other hand, the truth values of the
  \emph{even}-numbered variables is specified by the \emph{votes} of
  the voters $v_{3i}$, $1 \leq i \leq \ell$, which after these $\ell$
  deletions and after lexicographically reordering by name of the
  voter (as required by~$\cale$) will all have an even subscript. 

I rephrased it a bit.  There is not need for
  lexicographically reordering, since the voters are already
  ordered; for the full version, we really need to prove both
  directions separately and in more detail.}%
 It follows that the chair's goal can be reached by at most $k$ deletions of
  voters if and only if $(\exists b_1)\, (\forall b_2)\, \cdots
  (\exists b_{2\ell - 1})\, (\forall b_{2\ell})\, [F(x_1:=b_1,
  x_2:=b_2, \ldots , x_{2\ell}:=b_{2\ell}) \text{ evaluates to
    \emph{true}}]$, which is true if and only if $F \in \qbf'$.

  $\pspace$-hardness of $\onlineccav$ for the election system $\cale$
  defined above can be shown via essentially the same
  $\manyonereducesto$-reduction from $\qbf'$.  The only difference is
  that we now map the given $\qbf'$ instance $F$ to an instance of
  $\onlineccav$, which is defined exactly as the $\onlineccdv$
  instance constructed above, except that all voters $v_i$ with $i
  \equiv 0 \bmod 3$ are specified as registered voters, and all other
  voters are unregistered.  The correctness argument is analogous.

  The destructive cases can be shown analogously, by modifying the
  election system $\cale$ defined above as follows, yielding our
  modified system~$\cale'$: Whenever everyone loses (wins) in $\cale$,
  everyone wins (loses) in~$\cale'$.  It follows from
  Theorem~\ref{thm:general-pspace-upper-bound} and the above
  $\manyonereducesto$-reduction from $\qbf'$ that
  $\onlinesystemdcdv{\cale'}$ and $\onlinesystemdcav{\cale'}$ are both
  $\pspace$-complete.~\end{proofs}

For control by deleting or adding voters, the deletion or addition
limit $k$ is---both in the non-online case and in our 
online definition (which is what is used in 
Theorem~\ref{thm:general-pspace-lower-bound-deleting-adding-voters})---part 
of the problem instance.  
To better understand the source of the tremendous level of 
computational hardness 
Theorem~\ref{thm:general-pspace-lower-bound-deleting-adding-voters}
showed that these problems can have, let us now
consider
restrictions of these problems in which the deletion or addition 
limit is bounded by a
constant.  For a given election system $\cale$ and a fixed~$k$, let
$\onlinelimitccdv{k}$ be the restriction of $\onlineccdv$ to those
inputs whose deletion limit is at most~$k$, and define the problem
variant $\onlinelimitccavcale{k}$
analogously.  We will show in Theorem~\ref{thm:bounded-limit}
that this change in the definition---bounding
the deletion/addition bound---brings
the complexity of these problems from $\pspace$ down to~$\conp$.
(In contrast, limiting the number of 
\emph{candidates} to two was shown by 
Theorem~\ref{thm:general-pspace-lower-bound-deleting-adding-voters}
to leave these two problems PSPACE-complete.)

The $\conp$ upper bound follows immediately from the following theorem
about restricted polynomial-time alternating Turing machines,
which is of interest in its own right. If we define the weight 
of a path of an alternating Turing machine to be the number 
of 1's in the existential guesses along the path, 
what Theorem~\ref{thm:k-dour-QBF} says is 
that the class of languages accepted by polynomial-time 
alternating Turing machines whose accepting paths 
are weight-bounded is 
precisely coNP\@.
\EHnoteHandled{This theorem is new. We may want to mention it in the
introduction.}%

\begin{theorem}\label{thm:k-dour-QBF}
Let $k \geq 0$.
The class of languages accepted by polynomial-time alternating 
Turing machines 
that satisfy the property that 
on each accepting computation path the number of existential guesses on which 
the bit is guessed as 1 is at most $k$ 
is precisely $\conp$.
\EHnoteHandled{I deleted (from Lane's note)
``whose existential guess blocks each involve exactly
the guess of a single bit'' and I changed ``path'' to ``accepting path''
twice.}%
\end{theorem}

\begin{proofs}
We will show this by induction on $k$. 
It is immediate that the $k = 0$ case is precisely $\conp$.
To prove the inductive step, let $k > 0$ and
let $A$ be a language accepted by a 
polynomial-time alternating Turing machine
that satisfies the property that 
on each accepting computation path the number of existential guesses on which 
the bit is guessed as 1 is at most $k$.  
(That is, any path 
that contains at least $k+1$ guessed 1's in its existential 
guesses must have as its (leaf) value Reject rather than Accept.
Recall that each path of a 
polynomial-time alternating Turing machine 
has as its individual (leaf) value either Accept or Reject, and the 
overall action of the Turing machine is determined by the 
thought-experiment of 
applying the existential and universal node actions of the machine 
to those leaf values, resulting in an Accept or Reject at the root
that determines the machine's acceptance or rejection 
on the given input.)
We will show that 
$A$ is in $\conp$.

Throughout this proof, all $x_i$'s and $y_i$'s are over $\{0,1\}$, i.e.,
are bits.

Let $B$ be a polynomial-time computable ternary predicate 
and let $\ell(n)$ be a polynomial 
such that
for all $x$, $x \in A$ if and only if
\[\forall x_1 \exists y_1 \  \forall x_2 \exists y_2 \ 
\dots \forall x_{\ell(|x|)} \exists y_{\ell(|x|)} 
\left ( B(x,x_1\dots x_{\ell(|x|)},y_1 \dots y_{\ell(|x|)}) \wedge
\sum_{i=1}^{\ell(|x|)} y_i \leq k \right ).\]
Such a polynomial and predicate exist, 
since we can add extra quantifiers with dummy 
variables to make the quantifiers alternating and we can 
always guess an existentially-quantified dummy variable as 0.

We can rewrite the above as follows.
For all $x$, $x \in A$ if and only if
\[\small
\begin{array}{ccccc l}
\forall x_1 &
( \forall x_2 \exists y_2 & \forall x_3 \exists y_3 & \dots & 
\forall x_{\ell(|x|)} \exists y_{\ell(|x|)} &
\left ( B(x,x_1\dots x_{\ell(|x|)}, 1y_2 \dots y_{\ell(|x|)}) \wedge
\sum_{i=2}^{\ell(|x|)} y_i \leq k - 1 \right ) \vee{}  \\
& \forall x_2  & ( \forall x_3 \exists y_3 & \dots & 
\forall x_{\ell(|x|)} \exists y_{\ell(|x|)} &
\left( B(x,x_1\dots x_{\ell(|x|)}, 01y_3 \dots y_{\ell(|x|)}) \wedge
\sum_{i=3}^{\ell(|x|)} y_i \leq k - 1 \right ) \vee{} \\
&  & \forall x_3 & ( \dots & 
\forall x_{\ell(|x|)} \exists y_{\ell(|x|)} &
\left( B(x,x_1\dots x_{\ell(|x|)}, 001y_4 \dots y_{\ell(|x|)}) \wedge
\sum_{i=4}^{\ell(|x|)} y_i \leq k - 1 \right ) \vee{} \\
&&& \vdots &&\\
&  &  & & 
\forall x_{\ell(|x|)} &
\left( B(x,x_1\dots x_{\ell(|x|)}, 0^{{\ell(|x|)} -1} 1)
\wedge
\sum_{i={\ell(|x|)}+1}^{\ell(|x|)} y_i \leq k - 1 \right ) \vee{} \\
&  &  & &  &
B(x,x_1\dots x_{\ell(|x|)}, 0^{{\ell(|x|)}} ) \dots  ))). 
\end{array}
\]

(Of course, 
$\sum_{i={\ell(|x|)}+1}^{\ell(|x|)} y_i \leq k - 1$
is true, since $k>0$ in the present case.)
The long expression above 
is not quite in the right form to apply the inductive hypothesis.
In order to be able to do so, define language $C$ such that
$\pair{x, x_1 \dots x_r
} \in C$ if and only if $r \leq {\ell(|x|)}$ and
\[\forall x_{r+1} \exists y_{r+1} \dots \forall x_{\ell(|x|)} \exists y_{\ell(|x|)} 
\left( B(x,x_1\dots x_{\ell(|x|)}, 0^{r-1} 1 y_{r+1} \dots y_{\ell(|x|)}) \wedge
\sum_{i=r+1}^{\ell(|x|)} y_{i} \leq k - 1 \right ).\]
Clearly $C$ can be accepted by a 
polynomial-time alternating Turing machine
that satisfies the property that 
on each accepting computation path the number of existential guesses on which 
the bit is guessed as 1 is at most $k-1$. By the inductive hypothesis,
$C$ is in $\conp$.  Since $x \in A$ if and only if
\begin{eqnarray*}
\lefteqn{\forall x_1 (\pair{x,x_1
} \in C \vee
\forall x_2 ( \pair{x,x_1 x_2
} \in C \vee 
\forall x_3 ( \pair{x,x_1 x_2 x_3
} \in C \vee{}}\\
& & \dots \forall x_{\ell(|x|)} ( \pair{x,x_1 x_2 x_3 \dots x_{\ell(|x|)}
} \in C) \vee 
B(x,x_1\dots x_{\ell(|x|)}, 0^{{\ell(|x|)}}) \dots  ))),
\end{eqnarray*}
it follows 
that $A$ is in $\conp$.
(Why is it in $\conp$?  Note that its complement is in NP due to 
having a polynomial-length witnesses.  
Let $N$ be a fixed NP Turing machine accepting $\overline{C}$.
Our witness scheme for membership in $\overline{A}$ is:
Guess an
$x_1, \dots, x_{\ell(|x|)}$ 
such that $B(x,x_1\dots x_{\ell(|x|)}, 0^{{\ell(|x|)}})  $
holds and also guess
for each of 
$\pair{x,x_1
}$,  
$\pair{x,x_1 x_2
}$,~$\ldots$,  
$\pair{x,x_1 x_2 \dots x_{\ell(|x|)}
}$
an accepting path of $N$ on that input.)

\end{proofs}

\begin{theorem}
\label{thm:bounded-limit}
For each $k \geq 0$, the following hold:
\begin{enumerate}
\item (a)~For each election system $\cale$ with a polynomial-time winner
  problem, $\onlinelimitccdv{k}$ is in
  $\conp$.
(b)~There exists an election system $\cale$
with a polynomial-time
  winner problem such that $\onlinelimitccdv{k}$ is
$\conp$-complete, even when limited to 
  two candidates.

\item (a)~For each election system $\cale$ with a polynomial-time winner
  problem, $\onlinelimitccav{k}$ is in
  $\conp$.
(b)~There exists an election system $\cale$
with a polynomial-time
  winner problem such that 
  $\onlinelimitccav{k}$ 
 is
$\conp$-complete, even when limited to 
  two candidates.

\end{enumerate}
\end{theorem}

\sproofketch 
Parts 1(a) and 2(a) follow immediately from Theorem~\ref{thm:k-dour-QBF}.
\EHnoteHandled{Incorrect proof was:

For the (a) parts, for $k$ additions or deletions there are only at
most $\|V\|^k$ or $\binom{\|V\|}{k}$ ways to do those, so this can
easily be done by a polynomial-time disjunctive truth-table reduction
(see \cite{lad-lyn-sel:j:com} for the definition) to a $\conp$
problem, and so $\onlinelimitccdv{k}$ and $\onlinelimitccavcaleprime{k}$ are
easily seen to be in~$\conp$.

I do not think we need to mention anything about this (including why
this is incorrect) in the text).}%

Now consider part 1(b).  Even for $k=0$ (and in effect so for all~$k$, as
those have within them $k=0$ as subcases we can map to)
we claim that there is an election system $\cale$ with a polynomial-time
winner problem such that 
$\onlinelimitccdv{k}$ is easily shown to be $\conp$-hard, namely
\EHnoteHandled{We are not saying what $\cale$
and $\cale'$ are!  I am guessing that these are systems from
the proof of
Theorem~\ref{thm:general-pspace-lower-bound-deleting-adding-voters}.
But if so, we would only need
$\cale$ in this theorem.}%
by a
$\manyonereducesto$-reduction from the $\conp$-complete tautology
problem.  The mapping and $\cale$ are inspired by 
the proof of
Theorem~\ref{thm:general-pspace-lower-bound-deleting-adding-voters}:
We use the lexicographically least candidate name to be a proposed
tautology and we use the voters as tests of various assignments to it
(if the assignment satisfies, everyone wins).  So the problem can
force the chair's top choice (candidate~$a$, see the proof of
Theorem~\ref{thm:general-pspace-lower-bound-deleting-adding-voters})
to win exactly if the formula is a tautology.  As in 
the statement and proof
of Theorem~\ref{thm:general-pspace-lower-bound-deleting-adding-voters},
this reduction maps to outputs having only two candidates.

The proof sketch for part 2(b)
($\onlinelimitccav{k}$) is similar to that of part 1(b).  
The first (and
current) voter in our reduction is unregistered (but with $k=0$ she
obviously cannot be added), and the remaining voters are testing
assignments to a proposed tautology and we have only two candidates,
just as in the above proof sketch for $\onlinelimitccdv{k}$.~\eproof

\subsection{Control by Partition of Voters}

\begin{theorem}
\label{thm:general-pspace-lower-bound-partitioning-voters}  
There exist election systems $\cale$ and $\cale'$, whose winner
problems can be solved in polynomial time, such that $\onlineccpvtp$
and $\onlinesystemdcpvtp{\cale'}$ are $\pspace$-complete, even when
limited to two candidates.
\end{theorem}

\begin{proofs}
  This proof is similar in flavor to the proof of
  Theorem~\ref{thm:general-pspace-lower-bound-deleting-adding-voters},
  but since we now handle control by partition of voters, there are
  some
  decisive differences.

  The election system $\cale$ is now defined as follows.
\begin{algorithmusfall}
\item There is a candidate named RoundOne, and no voter is named
  Marker.  In this case, everyone loses.

\item There is a candidate named RoundOne and a voter named Marker.
  In this case, interpret---in our fixed, natural encoding---the
  lexicographically least candidate not named RoundOne as a boolean
  formula, $\Phi$, whose variable names must be the strings
   $x_1, x_2, \ldots ,
  x_{2\ell}$ for some~$\ell$, and $x_{2\ell}$ must actually appear in
  $\Phi$ (the others do not have to, but no variable other than $x_1,
  x_2, \ldots , x_{2\ell}$ can appear in~$\Phi$).  If this candidate
  is not of the required syntactic form, exactly RoundOne wins.  If
  the candidate set does not consist of exactly RoundOne and the above
  candidate, then exactly RoundOne wins.  If the voter list consists of
  exactly $2\ell + 1$ voters such that one voter is named Marker, one
  voter is named $v_1^{\mathrm{yes}}$ or $v_1^{\mathrm{no}}$, one
  voter is named $v_2$, one voter is named $v_3^{\mathrm{yes}}$ or
  $v_3^{\mathrm{no}}$, $\ldots$, one voter is named $v_{2\ell -
    1}^{\mathrm{yes}}$ or $v_{2\ell - 1}^{\mathrm{no}}$, and one voter
  is named $v_{2\ell}$, where all subscripts are given in binary, then
  assign the $2\ell$ variables of $\Phi$ as follows.  (If the voter
  list is not exactly that then exactly RoundOne wins.)  For each
  odd~$i$, $1 \leq i \leq 2\ell - 1$, set $x_i$ to \emph{true} if
  there is a voter named $v_i^{\mathrm{yes}}$ and to \emph{false} if
  there is a voter named $v_i^{\mathrm{no}}$.  For each even~$i$, $2
  \leq i \leq 2\ell$, set $x_i$ to \emph{true} if the voter named
  $v_i$ has the property that in her preference order RoundOne is the
  top choice, and otherwise set $x_i$ to \emph{false}.  If this
  assignment makes $\Phi$ \emph{true}, then the candidate not named
  RoundOne is the only winner, otherwise (exactly) RoundOne wins.
\ehnote{Changed things a bit because we have only two candidates.}%

\item There is no candidate named RoundOne.  In this case, everyone
  wins.
\end{algorithmusfall}
This ends the specification of~$\cale$.  Clearly, $\cale$ has a
polynomial-time winner problem, since it is just evaluating a fully
specified and assigned boolean formula, and doing various syntactic
checks.

Our online control by partition of voters problems are all in
$\pspace$ by Theorem~\ref{thm:general-pspace-upper-bound}.  To prove
$\pspace$-hardness, we again $\manyonereducesto$-reduce from the
$\pspace$-complete problem $\qbf'$ defined in the proof of
Theorem~\ref{thm:general-pspace-lower-bound-deleting-adding-voters}.
Let $F(x_1, \ldots, x_{2\ell})$ be a given $\qbf'$ instance, where
$x_{2\ell}$ actually occurs in~$F$. (If our input is syntactically
incorrect, then map it to a fixed nonmember of
our target problem.)
Our candidate set will be $C = \{\mathrm{RoundOne}, a\}$, where $a$
will in her name encode $F$ (without loss of
generality, that will not form the string ``RoundOne''),
$a$ will be our distinguished candidate,
our current voter will be $u = \tilde{v}_0$,
\JRnoteHandled{I renamed $\tilde{v}_1$ (Marker) to $\tilde{v}_0$ and shifted
  the subscripts of tilde voters below because in the voter/name
  scheme below the last voter is $\tilde{v}_{3\ell}$, not
  $\tilde{v}_{3\ell + 1}$.}%
\LAHnoteHandled{I didn't check that this renaming is fine, but 
I'll simply assume that it is, as Joerg noticed the need for it and 
did it.}%
the chair's preference order will be $a >_{\sigma} \mathrm{RoundOne}$,
and there will be  $3\ell + 1$ voters who vote in order
$\tilde{v}_0, \tilde{v}_1, \ldots , \tilde{v}_{3\ell}$,
where $\tilde{v}_0$ is named Marker,
and the remaining voters are named as follows:
\[
\begin{array}{@{}l ccc ccc c ccc@{}}
\text{voter} &
\tilde{v}_1 & \tilde{v}_2 & \tilde{v}_3 & 
\tilde{v}_4 & \tilde{v}_5 & \tilde{v}_6 & 
\cdots      & 
\tilde{v}_{3\ell - 2} & \tilde{v}_{3\ell - 1} & \tilde{v}_{3\ell} \\[1mm]
\text{name} &
v_1^{\mathrm{yes}} & v_1^{\mathrm{no}} & v_2 & 
v_3^{\mathrm{yes}} & v_3^{\mathrm{no}} & v_4 & 
\cdots      & 
v_{2\ell - 1}^{\mathrm{yes}} & v_{2\ell - 1}^{\mathrm{no}} & v_ {2\ell} 
\end{array}
\]
This ends our statement of the reduction.  Why does it work?

If $F \in \qbf'$, then
\begin{eqnarray}
\label{equ:qbf-prime}
 & & (\exists b_1)\, (\forall b_2)\, \cdots (\exists
b_{2\ell - 1})\, (\forall b_{2\ell}) \\ 
 & & [F(x_1:=b_1, x_2:=b_2, \ldots ,
x_{2\ell}:=b_{2\ell}) \text{ evaluates to \emph{true}}],
\nonumber
\end{eqnarray}
where the $b_i \in \{0,1\}$ are truth assignments.  So the partition
that puts Marker and all voters $v_i$, $i$ even,
\ehnote{We never put the even voters anywhere!}%
on one side, say into $V_{\mathrm{left}}$, and for
each $v_i^{\mathrm{yes}} / v_i^{\mathrm{no}}$ pair, $i$ odd, follows
(\ref{equ:qbf-prime}) by putting $v_i^{\mathrm{yes}}$ into
$V_{\mathrm{left}}$ and $v_i^{\mathrm{no}}$ into $V_{\mathrm{right}}$
if $b_i = 1$, and $v_i^{\mathrm{no}}$ into $V_{\mathrm{left}}$ and
$v_i^{\mathrm{yes}}$ into $V_{\mathrm{right}}$ if $b_i = 0$ (and
crucially note that the preference orders of the $v_i$, $i$ even, we
will have seen in the future can (in the future) effect
the future partition choices),
\jrnote{I wasn't sure if I could read everything in this sentence
correctly in the notes, please check.}%
\lahnote{i have not checked the JR note.}%
\ehnote{I guess it's ok now.}%
will by Case~2 have one first-round election (namely,
$(C,V_{\mathrm{left}})$) in which $a$ is the only
winner.
And in the other first-round election,
$(C,V_{\mathrm{right}})$, by Case~1 everyone, including RoundOne,
loses.  Thus,
only $a$ proceeds to the second-round
runoff election, where by Case~3 everyone wins, i.e., our
distinguished candidate $a$ wins.

In the other direction, suppose $F$ is syntactically correct, and it
is possible by some partition of voters to force ``$a$ or better'' (so
$a$) to be a winner.  Since RoundOne is in both first-round elections
(so Case~3 cannot occur), the only way candidate $a$ can be guaranteed
to even survive at least one first-round election is if we can
guarantee that Case~2 is satisfied.  But that means that $F \in \qbf'$.

Since our reduction can be computed in polynomial time, this shows
that $\onlineccpvtp$ is $\pspace$-hard.

\ehnote{This is ok for ECAI.  For the TR,  delete
the previous sentence.
For the full version, we may need to give more detail.}%
\jrnote{I deleted the previous sentence for the TR.}%
To show that $\onlinesystemdcpvtp{\cale'}$ is
$\pspace$-hard, we modify the election system $\cale$ defined above as
follows, yielding our modified system~$\cale'$: Most crucially, Case~2
of the election system description changes to now making everyone lose
if $\Phi$ evaluates to \emph{true} under the specified assignment, and
if $\Phi$ evaluates to \emph{false} (or there is any syntactic problem
regarding who is in the voter list) then everyone wins.
Case~3 changes to now having everyone lose, and Case~1 stays the same.
\ehnote{This was: `` and Case~1 does not need to
change, as it actually is (and even was in the description of~$\cale$)
irrelevant---for completeness of specification, one has to say what
happens in that case, but any well-specified outcome (such as ``all
win'' or ``all lose'') is equally good to make the proof work. ''

But that's nonsense.  If everyone wins in Case 1, then (assuming
RoundOne is a candidate), everyone will win by partition of voters
(since one of the partitions will not have voter Marker).  

I don't think Case 3 needs to change.  But it doesn't hurt.
And I think we may be able to ensure both winners or no winners,
so we may not need to change $\sigma$ either.
}%
The
$\manyonereducesto$-reduction from $\qbf'$ remains the same, except
that the chair's preference order will now be reversed to
$\mathrm{RoundOne} >_{\sigma} a$, and with these changes the reduction
can be shown to work correctly by arguments analogous to those in the
constructive case.~\end{proofs}

The above proof establishes
that there are election systems, with polynomial-time winner
problems, for which 
constructive
and destructive online control by partition of voters 
are $\pspace$-complete even when limited to 
two candidates.  Can we make do with one candidate and 
still have $\pspace$-hardness?  The following result
shows that if we could, then $\pspace$ would equal $\np \cap
\conp$.\footnote{Are elections with just one candidate 
even ever interesting in the real world?  We feel they 
sometimes are.
For example, a popular referendum---or for that 
matter a vote in a legislature on a bill---is 
essentially an up-or-down vote 
on one ``candidate.''  So is a vote on whether to recall 
an elected official, or to impeach a judge, or to ratify a person
who has been nominated for a sports hall of fame.}

\begin{theorem}
\label{thm:voter-partition-one-candidate}
\begin{enumerate}
\item For each election system $\cale$ with a polynomial-time winner
  problem, the problems $\onlineccpvtp$ and $\onlinedcpvtp$ when 
limited to one candidate are in~$\np$.

\item There exist election systems $\cale$ and $\cale'$ with
  polynomial-time winner problems such that the problems $\onlineccpvtp$ and
  $\onlinesystemdcpvtp{\cale'}$, even when restricted to one
  candidate, are $\np$-complete.
\jrnote{I added ``destructive'' to the statement of Thm.~52.1(b)
    and I integrated the two constructive claims from Thm.~60.1 here.}%
\lahnote{Adding destructive to the statement is correct and fixing 
a typo---I had left out the word but it is what is being proven there.
And what he means by 60.1 is just that he is merging 52.1 and 60.1 into
one statement---due to my first lahnote, we'll in the full probably
want to split them apart, so we can footnote just the constructive 
one with the NP-winner footnote... actually, that is a bit silly.
contrary to my earlier lane-note, let us not split them, but just 
(ONLY IN THE FULL VERSION) mention in the text that ``Part~1's claim
for the constructive case remains true even if the winner probably is 
merely assumed to belong to NP\.@''}%
\end{enumerate}
\end{theorem}

\begin{proofs}
  We give the proof for the destructive case.
\ehnote{Ok for ECAI.  For TR, replace by:
  ``We give the proof for the destructive case.''
For the full version, we need to add the proof.}%
\jrnote{I replaced this sentence by:
  ``We give the proof for the destructive case.'' for the TR.}%
  For the first part, with one candidate, $c$, every voter has the
  same preference as her full vote:~$c$.  So there is no sequentially
  revealed information, as in our model we know the voter names (and
  their order but here that does not matter) as part of our input.  So
  we just in $\np$ can guess every partition of the voters from~$u$,
  the current voter, onward, and see if one of those meets the chair's
  destructive goal, ``$c$ does not win.''

  For the second part, membership in $\np$ follows from the first
  part.  As to $\np$-hardness, let us $\manyonereducesto$-reduce from
  $\sat$.  The election system, $\cale'$, is defined as follows:
\begin{algorithmusfall}
\item If there are two or more candidates, everyone wins.

\item If there is one candidate and that candidate's name gives a
  syntactically correct boolean formula $\varphi$ that has, say, $k$
  variables, and there are exactly $k$ voters, and if we set the $i$th
  variable of $\varphi$ to \emph{true} exactly if $1$ is the lowest
  order bit of the voter whose name ranks $i$th in lexicographic order
  among the voters' names, then $\varphi$ is satisfied either by that
  assignment or by the bitwise complemented twin of that assignment,
  then everyone loses.

\item In all other cases (including syntactical problems), everyone
  wins.
\end{algorithmusfall}

The reduction $\sat \manyonereducesto \onlinesystemdcpvtp{\cale'}$ is
defined as follows.  Given a boolean formula $F(x_1, \ldots , x_k)$,
where without loss of generality all variables actually appear in~$F$,
we construct an $\onlinesystemdcpvtp{\cale'}$ instance with
candidate set $C = \{c\}$, where $c$ encodes~$F$, the voters are named
(in binary) $1, 2, \ldots , 2k$ and they vote in this order, $u=1$ is
the current voter, the distinguished candidate is~$c$, and the chair's
preference order $\sigma$ is~$c$.  Clearly, $c$ can be made not a
winner if and only if $F$ is satisfiable.  Why?

First, if $F$ is satisfiable then we can determine a satisfying
assignment by the partition choices we make among each voter pair
$(2i-1, 2i)$, $1 \leq i \leq k$, by choosing exactly one per pair for
the right-hand side of the partition, such that the left-hand side of
the partition has the bit-wise complement of that same satisfying
assignment.  So, by the definition of~$\cale'$, $c$ will not be a
winner in either first-round subelection, and so will not even be in
the final runoff election, which will have zero candidates, and so $c$
will not be a winner.

Second, if $c$ loses, by the election rule that proves that (Case~2 in
the definition of~$\cale'$), $F$ is satisfiable.

The constructive case can be shown analogously.~\end{proofs}

\begin{corollary}
\label{cor:voter-partition-one-candidate}
The following three statements are equivalent:
\begin{enumerate}
\item $\pspace = \np \cap \conp$.
\item There exists an election system $\cale$ with a polynomial-time
  winner problem such that $\onlinedcpvtp$ is $\pspace$-hard
  when restricted to one candidate.
\item There exists an election system $\cale$ with a polynomial-time
  winner problem such that $\onlineccpvtp$ is $\pspace$-hard 
  when restricted to one candidate.
\end{enumerate}
\end{corollary}

\begin{proofs}
  To show equivalence of the first two statements, suppose
  $\pspace = \np \cap \conp$.  So $\pspace = \np$.  The second
  statement now follows from the second part of
  Theorem~\ref{thm:voter-partition-one-candidate}.
  Conversely, by the second part's hypothesis and the first part of
  Theorem~\ref{thm:voter-partition-one-candidate}, we have $\pspace
  \subseteq \np$, which (since $\pspace = \rm coPSPACE$)
   is equivalent to $\pspace = \np \cap \conp$.
  The equivalence of the first and the third statements is proven
  analogously.~\end{proofs}
\lahnote{I removed the proof of this corollary as it is too
routine and trivial to include in the conference submission, 
though i guess it is fine in the full version we will send 
to journal.}%
\jrnote{I put the proof of this corollary back for the TR.}%

The analogues of the destructive cases of both parts of
Theorem~\ref{thm:voter-partition-one-candidate} also hold when
``online'' is removed, i.e., for the problem $\dcpvtp$.  
In contrast,
the \emph{constructive non-online} analogue of 
Theorem~\ref{thm:voter-partition-one-candidate}'s
first
part can be strengthened to a $\p$ upper bound.  
(Why can we get a $\p$ result here but not 
in Theorem~\ref{thm:voter-partition-one-candidate}?
The proof of the following 
result does not apply if some voters are already committed to
sides of the partition---it is assuming (and truly using the fact)
that we have full control of where \emph{all} voters go.  But in the
online setting, the current voter $u$ can be a voter who does
\emph{not} come first and so some voters may already be assigned to
sides of the partition.  And why do we get P for constructive but not
destructive?  The effect the following proof uses is 
specific to the constructive case.)
\jrnote{I have also omitted the discussion afterwards.
  And should Comment 59.2 still be added here?
  We have essentially made this point already in 
  Theorem~\ref{thm:voter-partition-one-candidate} and in the paragraph
  before Theorem~\ref{thm:nononline-voter-partition-one-candidate}.}%
\lahnote{I have rewritten the discussion 
before the theorem to include the key point regarding
this.  So the omitted discussion can be 
skipped permanently.}%
\begin{theorem}
\label{thm:nononline-voter-partition-one-candidate}
For each election system $\cale$ with a polynomial-time winner
problem, $\ccpvtp$, when restricted to 
one
candidate, is in~$\p$.
\end{theorem}

\begin{proofs}
  For the one candidate to win, she certainly must win the runoff, in
  which all voters vote.  Also, if she does win when all voters vote,
  then she can easily be made to survive the first round, using the
  partition structure $(V,\emptyset)$.  It follows from these two
  observations that constructive (non-online) control by partition of
  voters is possible if and only if the one candidate wins in the
  election with voter list~$V$.~\end{proofs}

\section{Online Control for Plurality}

We have seen in the previous section that online control can be very
hard, namely $\pspace$-complete, even for voting systems
whose winners can be determined in polynomial time.  In this section,
we study online control for plurality voting.  In this very simple
yet popular voting system, every voter gives one point to her most
preferred candidate, and all candidates with the most points win.  It
is known that non-online control by adding and by deleting voters can
be done in polynomial time, both in the
constructive
case (since the two relevant 
unique-winner-model
results of \cite{bar-tov-tri:j:control} as noted in 
\cite{fal-hem-hem:j:single-peaked-nearly}
also hold in the 
nonunique-winner model)
and in the
destructive case
(since 
we 
have checked and
here state as true that 
those unique-winner-model results
of~\cite{hem-hem-rot:j:destructive-control}
are easily seen to also hold in the 
nonunique-winner model).  
\LAHnoteHandled{The above handles the UW vs.\ NUW issue at THIS location, completely.
I have embedded into our source code, as invisible comments, the 
brief, very immediate proofs of both cases that are not previously 
stated in the literature.}%
We now
show that the corresponding types of online control are also easy.
\ehnote{Was: equally easy, but that may be too strong.}%
\begin{theorem}
\label{thm:plurality-deleting-adding}
The problems $\onlinesystemccdv{plurality}$, $\onlinesystemccav{plurality}$,
$\onlinesystemdcdv{plurality}$, and $\onlinesystemdcav{plurality}$
are in~$\p$.
\end{theorem}

\begin{proofs}
  For $\onlinesystemccdv{plurality}$, let $(C,u,V,\sigma,d)$ be a
  given basic OVCS, augmented by the additional information of online
  control by deleting voters: a deletion upper bound~$k$, for each
  voter $v$ before~$u$ a flag saying if $v$ was deleted and the vote
  cast by $v$ (if not deleted), where at most $k$ voters can be marked
  as deleted, and a vote to cast for $u$ (if $u$ is not to be
  deleted).  If $d$ is the chair's bottom choice in~$\sigma$, we are
  done, since the input then is trivially in
  $\onlinesystemccdv{plurality}$ (unless it is syntactically illegal).
  If exactly $k$ voters have been marked as already deleted, we can do
  no more deletions, so $u$ and all later voters go in, and we 
  assume (as this is the most challenging case) that all later voters
  vote for one particular candidate in
  $\Lambda_d = \{c \in C \condition c <_{\sigma} d\}$ that
  among the candidates in $\Lambda_d$ has the most
  first place votes after $u$ is put in,
\ehnote{Previous 17 words were missing}%
  and so we can
  easily answer the online control question.  If less than $k$ voters
  have been selected already for deletion, then delete $u$ if and only
  if $u$'s top choice is a highest scoring (with respect to the voters
  before~$u$) candidate in $\{c \in C \condition c <_\sigma d\}$.
  Then assume that all later voters vote for one particular candidate in
  $\Lambda_d = \{c \in C \condition c <_{\sigma} d\}$ that
  among the candidates in $\Lambda_d$ has the most
  first place votes after $u$ is put in.  And assume we delete as 
  many of those as the deletion amount left (after $u$) allows.
\jrnote{I've shifted the deletion argument a bit.}%
\lahnote{I have not checked the jrnote.}%
\ehnote{I don't think the rewrite is correct.  I've re-rewritten it.}%
It is easy to see whether this results in ``$d$ or
  better''
  \jrnote{I added ``or better.''}%
  \lahnote{I have not checked the jrnote but it sound as if that is correct
    and needed.}%
  \ehnote{Yes, that is correct and needed.}%
  being a winner (in which case our algorithm answers ``yes'') or not
  (in which case our algorithm answers ``no'').
(One might comment that it would suffice, 
especially to just handle the decision version, 
to follow the very simple ``operational'' approach mentioned 
on page \pageref{word:operational} of 
Section~\ref{s:motivation}.  However, we have given
a more dynamic description of the process both as we want to make 
clear how the chair can decide what action to take at each point
and as the description above is also helping establish the 
correctness of the actions taken.)
\LAHnoteHandled{Edith, you asked about what the Operational Approach
is referring to, and I have added a pageref into the text.}%

  For $\onlinesystemccav{plurality}$, let $(C,u,V,\sigma,d)$ be a
  given basic OVCS,
  augmented by the additional information of online control
  by adding voters: an addition upper bound $k$, for each voter the
  information of whether she is registered or not, and for each
  unregistered voter before $u$ the information of whether she has
  been added or not, the vote of each registered or added voter before
  $u$, and $u$'s potential vote.  Again, the question is trivial if $d$
  is the chair's bottom choice in~$\sigma$.  Otherwise, we can see what
  $u$'s vote is and if $k$ has yet been reached.  If $k$ has not been
  reached yet, we add $u$ if and only if $u$'s top choice belongs to
  $\{c \in C \condition c \geq_{\sigma}
  d\}$.\footnote{Sure enough, $u$'s top choice could be one of those
    candidates that end up having only few votes, so adding $u$ could
    be a wasted addition that will block some future good addition in
    some vote sequences, but in the worst case all future voters put
    first a candidate disliked by the chair; so our action is fine
    within the quantifier structure of the problem.}
\jrnote{I put this parenthetical into a footnote and slightly rephrase it,
please check.}%
\lahnote{I have not checked the jrnote.}%
\ehnote{It's fine.}%
  And in the worst case all future voters vote for the same member of
  $\{c \in C \condition c <_\sigma d\}$, which will be one that after
  $u$ votes has the most first-place votes among those.

  The two destructive cases
  can be handled analogously.  The main differences are, in both
  cases, that the question now is trivial to decide if $d$ is the
  chair's \emph{top} choice in~$\sigma$; in the deleting-voters case,
  that $u$ is to be deleted (provided the deletion limit $k$ has not
  been reached yet) if and only if $u$'s top choice is a highest
  scoring (with respect to the voters before~$u$) candidate in $\{c
  \in C \condition c \leq_{\sigma} d\}$; and in the adding-voters
  case, that $u$ is to be added (provided the addition limit $k$ has
  not been reached yet) if and only if $u$'s top choice belongs to
  $\{c \in C \condition c >_\sigma d\}$.  And, in both cases, we again
  assume that all future votes will belong to some particular member
  of $\{c \in C \condition c \leq_{\sigma} d\}$ that after $u$ votes
  has the most first-place votes among those candidates.~\end{proofs}

\OMIT{
\jrnote{Insert the following (perhaps as a footnote?) only if
Theorem~\ref{thm:plurality-partitioning-ties-eliminate} can be shown.}%
In many places in the literature tie-breaking models are treated as
not being of great importance.  However, there are some cases known
where the tie-breaking model does change problem
complexities---regarding tie-breaking related to partition, a contrast
between NP-completeness and membership in P was obtained 
in~\cite{hem-hem-rot:j:destructive-control}.
Our Theorems~\ref{thm:plurality-partitioning}
and~\ref{thm:plurality-partitioning-ties-eliminate} add two new such
cases: online constructive and destructive control by partition of
voters is $\conp$-hard in the model (ties-promote) we feel is most
natural and have adopted in this paper, but is in $\p$ in the other
most commonly studied partition tie-breaking model (which is known as
the \emph{ties-eliminate (TE)}
model~\cite{hem-hem-rot:j:destructive-control}, where only a unique
winner of a subelection proceeds to the runoff).
} %

Non-online control by partition of voters, in the model
we feel is most natural and have adopted in this paper (called 
``ties promote''), is 
$\np$-complete
in both the
constructive and destructive cases
(\cite{hem-hem-rot:j:destructive-control} showed this in
the unique-winner model, and we have checked and here state that 
NP-completeness also holds for the nonunique-winner model analogues).
\LAHnoteHandled{That is a UW citation.  But this paper is in the NUW 
model.  We need to check and then assert, here, that both those 
NP-completeness results also hold for the NUW model.  Perhaps 
that follows for free implicitly from the HHR proof, or perhaps
we'll have to tweak something.  Joerg, please check that both are 
still NP-complete in the NUW model, and then: (a) if it is easy
(or medium or even hard but routine), just 
send us an email with the easy proof, which we can embed as an
invisible comment in the source of the paper, as I did regarding 
DCDV and DCAV earlier in the paper; or (b) if it is interesting
enough, and hard, we can add it as an appendix B\@.  For now, I've 
assumed that (a) is the case, and have already rewritten the above 
body text to assume that you will do (a).}%
\EHnoteHandled{The UW vs.\ NUW issue at THIS location has been handled.
I have checked and embedded into our source code,
as invisible comments, Joerg's explanation 
of both cases that are not previously stated in the literature.}%
In contrast, the corresponding
types of online control are both $\conp$-hard.  This implies that
these problems cannot be in $\np$, unless $\np = \conp$, which is
considered to be highly unlikely.  It remains open whether or not they
are in $\conp$; we conjecture that they are not.
\ehnote{Conjecture is new.}%
\EHnoteHandled{I am not sure that I am still behind this conjecture.}
\LAHnoteHandled{After chatting with Edith: let's leave it in as a 
conjecture.  Even if people disprove it, that too is valuable and 
exciting.}%

\begin{theorem}
\label{thm:plurality-partitioning}
$\onlinesystemccpvtp{plurality}$ and $\onlinesystemdcpvtp{plurality}$
are $\conp$-hard.
\end{theorem}

\LAHnoteHandled{Edith, I'll hand back to you pages 41 and 42 of your marked
notes on the printout; please yourself type in your changes from 
those two pages, as they will be safer if typed in by you.}%
\EHnoteHandled{This was all minor and has been handled.}
\begin{proofs}
  We prove this by a reduction from the complement of the
  following $\np$-complete problem, Hitting Set: 
  Given a set $B = \{b_1, \ldots,
  b_m\}$, a nonempty collection ${\cal S} = \{S_1, \ldots, S_n\}$ of
  subsets of $B$, and a positive integer $k \leq m$, does ${\cal S}$
  have a hitting set of size at most~$k$, i.e., does there exist a set
  $B' \subseteq B$ such that $\|B'\| \leq k$ and for all $S_i \in
  {\cal S}$, $S_i \cap B' \neq \emptyset$.
\ehnote{Used to say:
 The reduction is reminiscent of the proof of $\np$-hardness of
 $\systemccpvtp{plurality}$ and $\systemdcpvtp{plurality}$
 from~\cite{hem-hem-rot:j:destructive-control}.

But it is pretty different.}%

We turn an instance  $(B,{\cal S},k)$ of Hitting Set
into the following instance of online partition of voters.
The set of candidates is
$\{c,w, b_1, \ldots, b_m\} \cup A$, where 
$A = \{a_i \ | \ 1 \leq i \leq 4mnk + 1\}$.
The current voter is~$u$.
The votes before $u$ that are on the left side of the
partition are exactly the same as the votes before $u$
that are on the right side of the partition.  Both sides of
the partition consists of the following votes.
\begin{itemize}
\item $4nk$ votes $c > w > \cdots$, where $\cdots$ denotes that the
remaining candidates follow in some arbitrary order.
\item $4nk$ votes $w > c > \cdots$.
\item For every $i$, $1 \leq i \leq n$,
$2k$ votes $S_i > c > \cdots$, where $S_i$ denotes the candidates
in $S_i$ in some arbitrary order.
\item For every $j$, $1 \leq j \leq m$, as many votes
$b_j > B - \{b_j\} > c > w > \cdots$ as needed to
make the score of $b_j$ equal to $4nk - 1$ in this subelection.
\jrnote{I added ``where $\score(b_j)$ is the number of
 points $b_j$ gets before $u$ casts her vote.''}%
\lahnote{I have not checked the jrnote.}%
\ehnote{That's not right; what matters is the score in the subelection.}%
\item For every $i$, $1 \leq i \leq 4mnk$, one vote
$a_i > c > \cdots$ and one vote $a_i > w > \cdots$.
\end{itemize}
\ehnote{I changed voter to vote in the subelections.}%
Voter $u$ votes $a_{4mnk+1} > w > \cdots$.  And there are $k$ voters after $u$.
The chair's top choice is $c$ and the chair's bottom choice is~$w$,
and the distinguished candidate is $c$ in the constructive case
(i.e., for $\onlinesystemccpvtp{plurality}$) and
$w$ in the destructive case (i.e., for $\onlinesystemdcpvtp{plurality}$).
\jrnote{I added the previous sentence.}%
\lahnote{I have not checked the jrnote.}%
\ehnote{It's ok}%

A simple but crucial observation is that no candidate $a \in A$ 
will ever make it to the final round, since her score
in the first round in either subelection will be at most $2 + k$,
which is less than $c$'s score in that subelection. 
If both $c$ and $w$ participate
in the final round, $c$ gains $8mnk$ points, $w$ gains 
$8mnk+1$ points, and no other candidate gains points from the voters
specified above
whose top choice was in $A$.

We will show that ${\cal S}$ does not have a hitting set of size at most $k$ if
and only if $c$ can always be made a winner in the constructed election, and
we will show that ${\cal S}$ does not have a hitting set of size $k$ if
and only if $w$ can always be made to not be a winner in the constructed
election.  This proves the theorem.

First suppose that ${\cal S}$ has a hitting set of size at most~$k$.
Let $B'$ be a hitting set of size~$k$.  $B'$ exists, since $k \leq m$.
Let the $k$ voters after $u$ vote such that the top choice
of the $i$th voter is the $i$th candidate in~$B'$.
Then, no matter how we partition the voters, the set
of candidates that participate in the final round
is $\{c,w\} \cup B'$.  The scores in the final round are
as follows:
(a)~$\score(c) = 8nk + 8mnk$,
(b)~$\score(w) = 8nk + 8mnk + 1$, and
(c)~$\sum_{b \in B'} \score(b) = 8mnk - 2m + k$.
It follows that $c$ is not a winner and that $w$ is a winner.

For the converse, suppose that 
${\cal S}$ does not have a hitting set of size at most $k$.
Partition by putting $u$ and all voters after $u$ in the same
first-round election.  Then the set of candidates in the final
round is $\{c,w\} \cup B'$, where $B' \subseteq B$ and $\|B'\| \leq k$.
Since $B'$ is not a hitting set,
in the final round
$c$ gains at least $4k$ points from voters voting
$S_i > c > \cdots$ such that $S_i \cap B' = \emptyset$.
Thus in the final election the following hold:
(a)~$\score(c) \geq 8nk + 8mnk + 4k$,
(b)~$\score(w) \leq 8nk + 8mnk + 1 + k$, and
(c)~$\sum_{b \in B'} \score(b) \leq 8mnk - 2m + k$.
\LAHnoteHandled{
Edith's notes suggest that (c) should 
be changed to:
``(c)~for all $b \in B'$,  $\score(b) \leq 8mnk - 2m + k$.''
I have NOT done that as this is part of a 2-page segment of 
proof editing that Edith will be doing directly as the pages are 
so crowded with her edits that it is safer for her to do them
herself.}%
\EHnoteHandled{This was minor and is now handled.}%
It follows that $c$ is the unique winner of this election.~\end{proofs}

\OMIT{
\begin{theorem}
\label{thm:plurality-partitioning-ties-eliminate}
$\onlinesystemccpvte{plurality}$ and $\onlinesystemdcpvte{plurality}$
are both in~$\p$.
\end{theorem}
} %

\section{Conclusions and Open Questions}

Inspired by the maxi-min approach of online algorithms, we studied
online voter control in sequential voting.  We showed that for
suitably constructed election systems with polynomial-time winner
problems, the resulting voter-control problems can be extremely hard, namely
$\pspace$-complete, even for just two candidates.  
We additionally obtain $\conp$-completeness for the 
deleting/adding-voter cases, even for just two candidates, 
when there is a bounded deletion/addition
limit.  
For plurality, things are easier still:
Online control by deleting or adding voters is in polynomial time for
plurality, just as
in the non-online case.  

Attractive future directions include the study of additional
natural election systems.  
Can one obtain PSPACE-completeness results for highly natural, existing
systems, for example?
Another interesting
direction would be to investigate online control 
through a typical-case analysis of heuristic 
approaches (such as, for example,
\cite{mcc-pri-sli:j:approximability-of-dodgson,hem-hom:j:dodgson-greedy}
do rigorously in a winner-problem setting, see 
also~\cite{rot-sch:j:typical-case-challenges}).
\jrnote{Should/can we mention the ECAI short paper and the AAAI paper
  here? }%
\lahnote{ANSWER:  I've cited manipulation from elsewhere in the paper!
and the ECAI version should not cite the ECAI short paper, but the 
TR version should say something like:
This paper studies online voter control, and a sister paper studies online candidate control~backslash-cite$\{$the arxiv TR$\}$.
}%
\jrnote{For the TR I replaced the last sentence to point to the sister
 paper on online candidate control.}%

\section*{Acknowledgments}
Parts of this paper appeared in preliminary form in 
ECAI-2012~\cite{hem-hem-rot:c:online-voter-control}.
We are very grateful to the 
referees
for their comments and suggestions.

\bibliographystyle{alpha}

\begin{thebibliography}{FHHR09b}

\bibitem[BE98]{bor-ely:b:online-algorithms}
A.~Borodin and R.~{El-Yaniv}.
\newblock {\em Online Computation and Competitive Analysis}.
\newblock Cambridge University Press, 1998.

\bibitem[BH08]{buh-hit:c:np-hard-sets-and-density}
H.~Buhrman and J.~Hitchcock.
\newblock {NP}-hard sets are exponentially dense unless
  {coNP}$\,\subseteq\,${NP}/poly.
\newblock In {\em Proceedings of the 23rd Annual IEEE Conference on
  Computational Complexity}, pages 1--7. IEEE Computer Society Press, June
  2008.

\bibitem[BR16]{bau-rot:b:preference-aggregation-by-voting}
D.~Baumeister and J.~Rothe.
\newblock Preference aggregation by voting.
\newblock In J.~Rothe, editor, {\em Economics and Computation: {An}
  Introduction to Algorithmic Game Theory, Computational Social Choice, and
  Fair Division}, pages 197--325. Springer, 2016.

\bibitem[BTT92]{bar-tov-tri:j:control}
J.~{{Bartholdi}}, III, C.~Tovey, and M.~Trick.
\newblock How hard is it to control an election?
\newblock {\em Mathematical and Computer Modeling}, 16(8/9):27--40, 1992.

\bibitem[CCHO05]{cai-cha-hem-ogi:j:s2-and-lowness}
J.~Cai, V.~Chakaravarthy, L.~Hemaspaandra, and M.~Ogihara.
\newblock Competing provers yield improved {Karp--Lipton} collapse results.
\newblock {\em Information and Computation}, 198(1):1--23, 2005.

\bibitem[CKS81]{cha-koz-sto:j:alternation}
A.~Chandra, D.~Kozen, and L.~Stockmeyer.
\newblock Alternation.
\newblock {\em Journal of the ACM}, 26(1):114--133, 1981.

\bibitem[DE10]{des-elk:c:sequential-voting}
Y.~Desmedt and E.~Elkind.
\newblock Equilibria of plurality voting with abstentions.
\newblock In {\em Proceedings of the 11th ACM Conference on Electronic
  Commerce}, pages 347--356. ACM Press, June 2010.

\bibitem[DKNS01]{dwo-kum-nao-siv:c:rank-aggregation}
C.~Dwork, R.~Kumar, M.~Naor, and D.~Sivakumar.
\newblock Rank aggregation methods for the web.
\newblock In {\em Proceedings of the 10th International World Wide Web
  Conference}, pages 613--622. ACM Press, March 2001.

\bibitem[DP01]{dek-pic:j:sequential-voting-binary-elections}
E.~Dekel and M.~Piccione.
\newblock Sequential voting procedures in symmetric binary elections.
\newblock {\em Journal of Political Economy}, 108(1):34--55, 2001.

\bibitem[EFRS15]{erd-fel-rot-sch:j:control-in-bucklin-and-fallback-voting}
G.~Erd\'{e}lyi, M.~Fellows, J.~Rothe, and L.~Schend.
\newblock Control complexity in {Bucklin} and fallback voting: {A} theoretical
  analysis.
\newblock {\em Journal of Computer and System Sciences}, 81(4):632--660, 2015.

\bibitem[ENR09]{erd-now-rot:j:sp-av}
G.~Erd\'{e}lyi, M.~Nowak, and J.~Rothe.
\newblock Sincere-strategy preference-based approval voting fully resists
  constructive control and broadly resists destructive control.
\newblock {\em Mathematical Logic Quarterly}, 55(4):425--443, 2009.

\bibitem[ER97]{eph-ros:j:multiagent-planning}
E.~Ephrati and J.~Rosenschein.
\newblock A heuristic technique for multi-agent planning.
\newblock {\em Annals of Mathematics and Artificial Intelligence},
  20(1--4):13--67, 1997.

\bibitem[FHH10]{fal-hem-hem:j:cacm-survey}
P.~Faliszewski, E.~Hemaspaandra, and L.~Hemaspaandra.
\newblock Using complexity to protect elections.
\newblock {\em Communications of the ACM}, 53(11):74--82, 2010.

\bibitem[FHH13]{fit-hem-hem:c:control-manipulation}
Z.~Fitzsimmons, E.~Hemaspaandra, and L.~Hemaspaandra.
\newblock Control in the presence of manipulators: {Cooperative} and
  competitive cases.
\newblock In {\em Proceedings of the 23rd International Joint Conference on
  Artificial Intelligence}, pages 113--119. AAAI Press, August 2013.

\bibitem[FHH14]{fal-hem-hem:j:single-peaked-nearly}
P.~Faliszewski, E.~Hemaspaandra, and L.~Hemaspaandra.
\newblock The complexity of manipulative attacks in nearly single-peaked
  electorates.
\newblock {\em Artificial Intelligence}, 207:69--99, 2014.

\bibitem[FHHR09a]{fal-hem-hem-rot:j:llull}
P.~Faliszewski, E.~Hemaspaandra, L.~Hemaspaandra, and J.~Rothe.
\newblock Llull and {Copeland} voting computationally resist bribery and
  constructive control.
\newblock {\em Journal of Artificial Intelligence Research}, 35:275--341, 2009.

\bibitem[FHHR09b]{fal-hem-hem-rot:b:richer}
P.~Faliszewski, E.~Hemaspaandra, L.~Hemaspaandra, and J.~Rothe.
\newblock A richer understanding of the complexity of election systems.
\newblock In S.~Ravi and S.~Shukla, editors, {\em Fundamental Problems in
  Computing: {Essays} in Honor of {Professor} {Daniel} {J.} {Rosenkrantz}},
  pages 375--406. Springer, 2009.

\bibitem[FR16]{fal-rot:b:handbook-comsoc-control-and-bribery}
P.~Faliszewski and J.~Rothe.
\newblock Control and bribery in voting.
\newblock In F.~Brandt, V.~Conitzer, U.~Endriss, J.~Lang, and A.~Procaccia,
  editors, {\em Handbook of Computational Social Choice}, pages 146--168.
  Cambridge University Press, 2016.

\bibitem[GMHS99]{gho-her-mun-sen:c:voting-for-movies}
S.~Ghosh, M.~Mundhe, K.~Hernandez, and S.~Sen.
\newblock Voting for movies: {The} anatomy of recommender systems.
\newblock In {\em Proceedings of the 3rd Annual Conference on Autonomous
  Agents}, pages 434--435. ACM Press, 1999.

\bibitem[HH09]{hem-hom:j:dodgson-greedy}
C.~Homan and L.~Hemaspaandra.
\newblock Guarantees for the success frequency of an algorithm for finding
  {Dodgson}-election winners.
\newblock {\em Journal of Heuristics}, 15(4):403--423, 2009.

\bibitem[HHR07]{hem-hem-rot:j:destructive-control}
E.~Hemaspaandra, L.~Hemaspaandra, and J.~Rothe.
\newblock Anyone but him: {The} complexity of precluding an alternative.
\newblock {\em Artificial Intelligence}, 171(5--6):255--285, 2007.

\bibitem[HHR09]{hem-hem-rot:j:hybrid}
E.~Hemaspaandra, L.~Hemaspaandra, and J.~Rothe.
\newblock Hybrid elections broaden complexity-theoretic resistance to control.
\newblock {\em Mathematical Logic Quarterly}, 55(4):397--424, 2009.

\bibitem[HHR12a]{hem-hem-rot:c:online-candidate-control}
E.~Hemaspaandra, L.~Hemaspaandra, and J.~Rothe.
\newblock Controlling candidate-sequential elections.
\newblock In {\em Proceedings of the 20th European Conference on Artificial
  Intelligence}, pages 905--906. IOS Press, August 2012.

\bibitem[HHR12b]{hem-hem-rot:c:online-voter-control}
E.~Hemaspaandra, L.~Hemaspaandra, and J.~Rothe.
\newblock Online voter control in sequential elections.
\newblock In {\em Proceedings of the 20th European Conference on Artificial
  Intelligence}, pages 396--401. IOS Press, August 2012.

\bibitem[HHR13]{hem-hem-rot:cOutByJour:online-manipulation}
E.~Hemaspaandra, L.~Hemaspaandra, and J.~Rothe.
\newblock The complexity of online manipulation of sequential elections.
\newblock In {\em Proceedings of the 14th Conference on Theoretical Aspects of
  Rationality and Knowledge}, pages 111--120. \mbox{TARK.org}, January 2013.

\bibitem[HHR14]{hem-hem-rot:j:online-manipulation}
E.~Hemaspaandra, L.~Hemaspaandra, and J.~Rothe.
\newblock The complexity of online manipulation of sequential elections.
\newblock {\em Journal of Computer and System Sciences}, 80(4):697--710, 2014.

\bibitem[HHR16]{hem-hem-rot:b:single-peaked-chapter}
E.~Hemaspaandra, L.~Hemaspaandra, and J.~Rothe.
\newblock The complexity of manipulative actions in single-peaked societies.
\newblock In J.~Rothe, editor, {\em Economics and Computation: {An}
  Introduction to Algorithmic Game Theory, Computational Social Choice, and
  Fair Division}, pages 327--360. Springer, 2016.

\bibitem[HU79]{hop-ull:b:automata}
J.~Hopcroft and J.~Ullman.
\newblock {\em Introduction to Automata Theory, Languages, and Computation}.
\newblock Addison-Wesley, 1979.

\bibitem[HW12]{hem-wil:j:heuristic-algorithms-correctness-frequency}
L.~Hemaspaandra and R.~Williams.
\newblock An atypical survey of typical-case heuristic algorithms.
\newblock {\em SIGACT News}, 43(4):71--89, 2012.

\bibitem[MPS08]{mcc-pri-sli:j:approximability-of-dodgson}
J.~{McCabe-Dansted}, G.~Pritchard, and A.~Slinko.
\newblock Approximability of {Dodgson's} rule.
\newblock {\em Social Choice and Welfare}, 31(2):311--330, 2008.

\bibitem[OL14]{luc-ore:c:budgeted-social-choice}
J.~Oren and B.~Lucier.
\newblock Online (budgeted) social choice.
\newblock In {\em Proceedings of the 28th AAAI Conference on Artificial
  Intelligence}, pages 1456--1462. AAAI Press, July 2014.

\bibitem[Pap94]{pap:b:complexity}
C.~Papadimitriou.
\newblock {\em Computational Complexity}.
\newblock Addison-Wesley, 1994.

\bibitem[PP13]{par-pro:c:dynamic-social-choice}
D.~Parkes and A.~Procaccia.
\newblock Dynamic social choice with evolving preferences.
\newblock In {\em Proceedings of the 27th AAAI Conference on Artificial
  Intelligence}, pages 767--773. AAAI Press, July 2013.

\bibitem[RS13]{rot-sch:j:typical-case-challenges}
J.~Rothe and L.~Schend.
\newblock Challenges to complexity shields that are supposed to protect
  elections against manipulation and control: {A} survey.
\newblock {\em Annals of Mathematics and Artificial Intelligence},
  68(1--3):161--193, 2013.

\bibitem[Sim69]{sim:b:bounded-rationality}
H.~Simon.
\newblock {\em The Sciences of the Artificial}.
\newblock MIT Press, 1969.
\newblock Third edition, 1996.

\bibitem[Slo93]{slo:j:sequential-voting}
B.~Sloth.
\newblock The theory of voting and equilibria in noncooperative games.
\newblock {\em Games and Economic Behavior}, 5(1):152--169, 1993.

\bibitem[Ten04]{ten:c:transitive-voting}
M.~Tennenholtz.
\newblock Transitive voting.
\newblock In {\em Proceedings of the 5th ACM Conference on Electronic
  Commerce}, pages 230--231. ACM Press, July 2004.

\bibitem[XC10]{con-xia:c:stackelberg-sequential}
L.~Xia and V.~Conitzer.
\newblock Stackelberg voting games: {Computational} aspects and paradoxes.
\newblock In {\em Proceedings of the 24th AAAI Conference on Artificial
  Intelligence}, pages 697--702. AAAI Press, July 2010.

\end{thebibliography}
\end{document}